\documentclass[11pt]{article}
\usepackage{amsmath,amssymb,latexsym,epsfig,multirow,graphicx,longtable,rotating, hyperref, xcolor}
\usepackage[compress]{cite}
\usepackage{caption}
\usepackage{subcaption}
\usepackage[normalem]{ulem}

\newcommand{\al}{\alpha}
\newcommand{\ba}{\beta}

\newcommand{\veps}{\varepsilon}

\newcommand{\ka}{\kappa}

\newcommand{\inblue}[1]{\textcolor{blue}{#1}}

\begin{document}

\title{A Cartesian FMM-accelerated Galerkin boundary integral \\
Poisson-Boltzmann solver}

\author{Jiahui~Chen\inblue{$^1$},~Johannes Tausch\inblue{$^2$},~Weihua~Geng{$^2$}\footnote{
Corresponding author. Tel: 1-214-7682252, Fax: 1-214-7682355,
Email: wgeng@smu.edu}\\
\\
{$^1$\small \it     Department of Mathematics,
Michigan State University, East Lansing, MI 48824, USA}\\
$^2$\small \it     Department of Mathematics,
Southern Methodist University, Dallas, TX 75275, USA\\
}

\date{\today}
\maketitle

\begin{abstract}
\indent 
The Poisson-Boltzmann model is an effective and popular approach 
for modeling solvated biomolecules in continuum solvent with dissolved electrolytes. 
In this paper, we report our recent work in developing a Galerkin boundary integral method 
for solving the Poisson-Boltzmann (PB) equation. 
The solver has combined advantages in accuracy, efficiency, and memory usage
as it applies a well-posed boundary integral formulation 
to circumvent many numerical difficulties associated with the PB equation
and 
uses an $O(N)$ Cartesian Fast Multipole Method (FMM)  to accelerate the GMRES iteration. 
In addition, special numerical treatments such as
adaptive FMM order,
block diagonal preconditioners, 
Galerkin discretization,  
and Duffy's transformation are combined to improve the performance of the solver,
which is validated on benchmark Kirkwood's sphere and a series of testing proteins.

\vspace*{1cm}


\end{abstract}

\newpage

\section{Introduction}
In biomolecular simulations, electrostatic interactions are of paramount importance 
due to their ubiquitous existence and significant contribution in the force fields, 
which governs the dynamics of molecular simulation.  
However, computing non-bonded interactions is challenging 
since these pairwise interactions are long-range with $O(N^2)$ computational cost, 
which could be prohibitively expensive for large systems. 
To reduce the degree of freedom of the system in terms of electrostatic interactions,  
an implicit solvent Poisson-Boltzmann (PB) model is used \cite{Baker:2005}. 
In this model, the explicit water molecules are treated as continuum 
and the dissolved electrolytes are approximated using the statistical Boltzmann distribution. 
The PB model has broad applications in biomolecular simulations such as 
protein structure \cite{Cherezov:2007}, 
protein-protein interaction \cite{Huang:2012}, 
chromatin packing \cite{Beard:2001}, 
pKa \cite{Alexov:2011,Hu:2018,Chen:2021a}, 
membrane \cite{Zhou:2010}, 
binding energy \cite{Nguyen:2017}, 
solvation free energy\cite{Wagoner:2006},  
ion channel profiling \cite{Unwin:2005}, etc.

The PB model is an elliptic interface problem 
with several numerical difficulties 
such as 
discontinuous dielectric coefficients, 
singular sources, 
a complex interface, 
and unbounded domains. 
Grid-based finite difference or finite volume discretization that discretize the entire
volumetric domain have been developed in, e.g.,
\cite{
Baker:2004, 
Im:1998, 
Honig:1995,
Baker:2001,
Luo:2002,
Deng:2015,
Ying:2015}.
The grid-based discretization is efficient and robust and is 
therefore popular.  
However, solvers that are based on discretizing the partial
differential equation may suffer from accuracy reduction due to 
discontinuity of the coefficients, 
non-smoothness of the solution, 
singularity of the sources, 
and truncation of the domains, 
unless special interface \cite{Qiao:2006, Yu:2007} 
and singularity \cite{Geng:2007, Cai:2009,Geng:2017, Lee:2020} treatments are applied. 
These treatments come at the price of more complicated discretization scheme 
and possibly reduced convergence speed of the iterative solver for the
linear system.

An alternative approach is to reformulate the PB equation as a
boundary integral equation and use the boundary elements 
to discretize the molecular surface,
e.g.~\cite{
Juffer:1991,
Boschitsch:2002,
Lu:2007,
Greengard:2009,
Bajaj:2011,
Zhang:2013,
Geng:2013b,
Zhong:2018,
Quan:2019}.
Besides the reduction from
three dimensional space to the two dimensional molecular boundary,
this approach has the advantage that singular charges, interface
conditions, and far-field condition are incorporated analytically in
the formulation, and hence do not impose additional approximation
errors.

In addition, due to the structures hidden in the linear algebraic system 
after the discretization of the boundary integral and molecular surface, 
the matrix-vector product in each iteration can be accelerated by fast methods 
such as fast multipole methods (FMM) \cite{Greengard:2002,Tausch:2004,
Boschitsch:2002,Lu:2007,Bajaj:2011}
and
treecodes~\cite{Barnes:1986, Li:2009, Wang:2020a}.
Our recently developed treecode-accelerated boundary integral (TABI) Poisson-Boltzmann solver 
\cite{Geng:2013b} is an example of a code that combines the advantages of 
both boundary integral equation and multipole methods. 
The TABI solver uses the well-posed derivative form of the Fredholm second kind integral equation \cite{Juffer:1991}
and the $O(N\log N)$ treecode \cite{Li:2009}. 
It also has advantages in memory use and parallelization
\cite{Geng:2013b, Chen:2021}. The TABI solver has been
used by many computational biophysics/biochemistry groups and it has
been disseminated as a standalone code \cite{Geng:2013b} or as a contributing module of the popular APBS software package \cite{Baker:2001a,Jurrus:2018}.

Recently, based on feedback from TABI solver users 
and our gained experience in theoretical development and practical applications, 
we realized that we could still improve the TABI solver in the following aspects.
First, the $O(N \log N)$ treecode can be replaced by the $O(N)$ FMM method 
with manageable extra costs in memory usage and complexity of the algorithms. 
Second, the singularity that occurs 
when the Poisson's or Yukawa's kernel is evaluated 
was previously handled by simply removing the singular triangle \cite{Geng:2013b, Lu:2007}
in fact can be treated by using the Duffy transformation \cite{Duffy:1982} analytically, 
achieving improved accuracy. 
Third,  the collocation scheme used in TABI solver 
can be updated by using Galerkin discretization
with further advantage in maintaining desired accuracy. 
Fourth, the treecode-based preconditioning scheme that was used
in TABI solver \cite{Chen:2018a} can be similarly developed and used under the FMM frame, 
receiving significant improvement in convergence and robustness. 
By combining all these new features, we developed a Cartesian fast multipole method (FMM) 
accelerated Galerkin boundary integral (FAGBI) Poisson--Boltzmann
solver. In the remainder of this article, we provide more detail
about the theoretical background of the numerical algorithms related
to the FAGBI solver. We conclude with a discussion of the numerical
results obtained with our implementation.

\section{Theory and Algorithms}
\label{theory}

In this section we briefly describe 
the Poisson-Boltzmann (PB) implicit solvent model and review the
boundary integral form of the PB equation and its
Galerkin discretization. 
Based on this background information, 
we then provide details of our recently developed FMM-accelerated Galerkin
boundary integral (FAGBI)  Poisson-Boltzmann solver, which involves
the boundary integral form, multipole expansion scheme, and a  block diagonal preconditioning scheme.

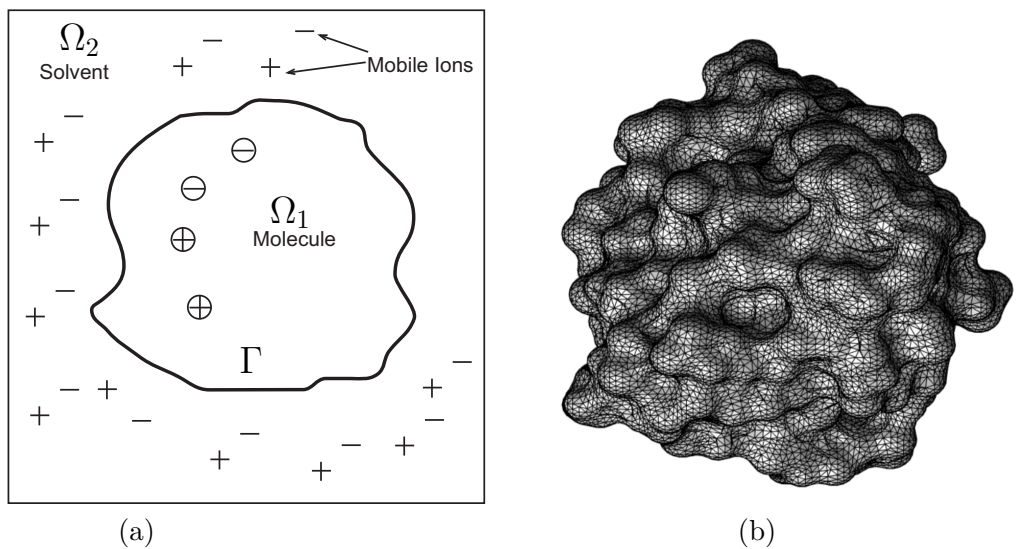
\begin{figure}[htb]
\setlength{\unitlength}{1cm}
\begin{picture}(5,1)
\put( 5.2, -3.2){\Large $\Omega_1$}
\put( 2.4, -0.9){\Large $\Omega_2$}
\put( 4.8, -5.2){\Large $\Gamma$}
\end{picture}
\begin{center}
\includegraphics[width=2.5in]{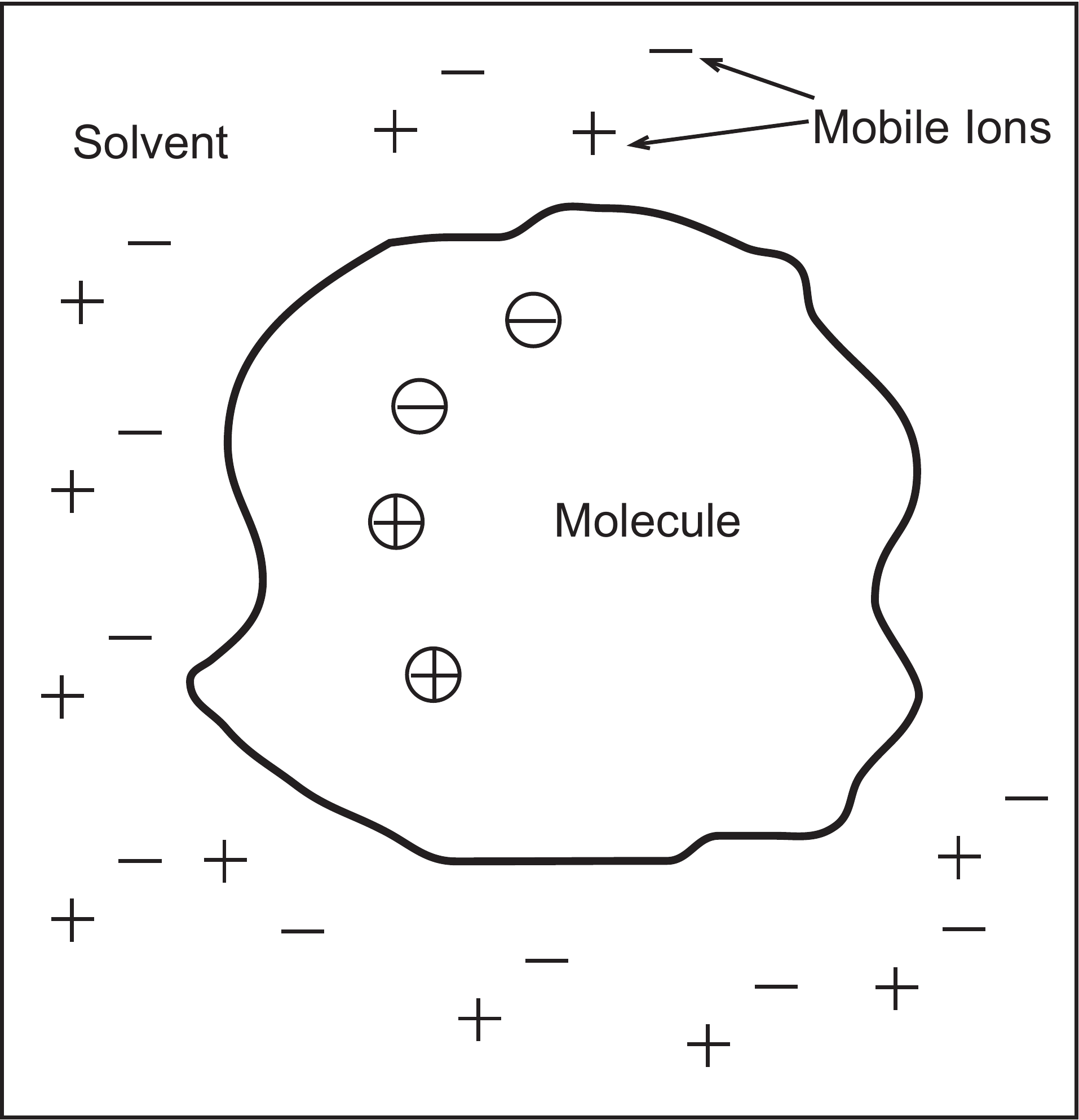}~~~~~
\includegraphics[width=2.5in]{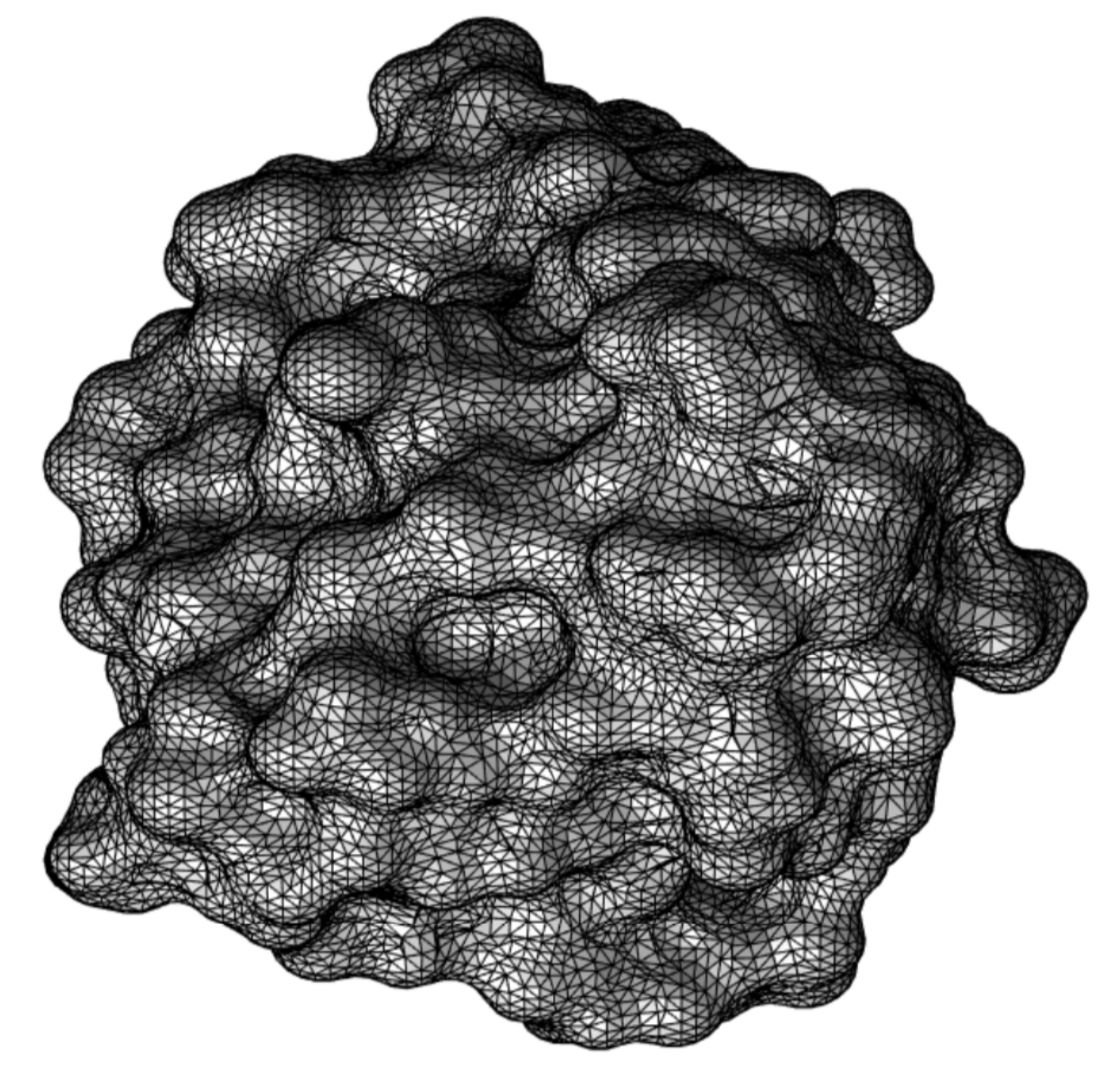}\\
\hskip -0.7in (a) \hskip 3in (b)
\caption{
Schematic models;
(a) the PB implicit solvent model, in which
the molecular surface $\Gamma$ separates space into the solute region $\Omega_1$ 
and solvent region $\Omega_2$; 
(b) the triangulation of molecular surface of protein Barstar at MSMS \cite{Sanner:1996} density $d=5$ (\# of vertices per \AA$^2$). 
}
\label{fig_1}
\end{center}
\end{figure}

\subsection{The Poisson-Boltzmann (PB) model for a solvated biomolecule}

The PB model for a solvated biomolecule is depicted in Fig.~\ref{fig_1}(a) in which
the molecular surface $\Gamma$ 
separates the solute domain $\Omega_1$ from the solvent domain $\Omega_2$.
Figure~\ref{fig_1}(b) is an example of the molecular surface $\Gamma$ 
as the triangulated surface of protein barstar \cite{Dong:2003}.  
In domain $\Omega_1$, 
the solute is represented by $N_c$ partial charges $q_k$ located at atomic centers ${\bf r}_k$ for $k=1,\cdots,N_c$,
while in domain $\Omega_2$,
a distribution of ions is described by a Boltzmann distribution 
and we consider a linearized version in this study. 
The solute domain has a low dielectric constant $\epsilon_1$
and
the solvent domain has a high dielectric constant $\epsilon_2$.
The modified inverse Debye length $\bar{\kappa}$ is given as $\bar{\kappa}^2 = \epsilon_2\kappa^2$,
where $\kappa$ is the inverse Debye length measuring the ionic strength;
$\bar{\kappa} = 0$ in $\Omega_1$ and is nonzero only in $\Omega_2$. 
The electrostatic potential $\phi({\bf x})$ satisfies the linear PB equation,
\begin{equation}
-\nabla \cdot \epsilon({\bf x}) \nabla \phi({\bf x}) + \bar{\kappa}^2({\bf x})\phi({\bf x}) =
\sum_{k=1}^{N_c} q_k \delta({\bf x}-{\bf x}_k),
\label{eqNPBE}
\end{equation}
subject to continuity conditions for the potential and electric flux density on $\Gamma$,
\begin{equation}
[\phi] = 0, \quad [\epsilon \phi_\nu] = 0,
\label{eqJump}
\end{equation}
where 
$[f] = f_1 - f_2$ is the difference of the quantity $f$ across the interface,
and
$\phi_\nu = \partial\phi/\partial{\nu}$ is the partial derivative in the outward normal direction $\nu$.
The model also incorporates the far-field condition,
\begin{equation}
\label{eqfar-field}
\lim_{{\bf x} \rightarrow \infty} \phi({\bf x }) = 0.
\end{equation}
Note that Eqs.~(\ref{eqNPBE})-(\ref{eqfar-field}) 
define a boundary value problem for the potential $\phi({\bf x})$
which in general must be solved numerically.

\subsection{Boundary integral form of PB model}

This section summarizes the well-conditioned boundary integral form of the 
PB implicit solvent model we employ~\cite{Juffer:1991,Geng:2013b}.
Applying Green's second identity and properties of fundamental solutions 
to Eq.~(\ref{eqNPBE}) yields the electrostatic potential in each domain,
\begin{subequations}
\begin{align}
\label{eqbim_1}
\phi({\bf x}) = 
&\int_\Gamma \left[G_0({\bf x},
{\bf y})\frac{\partial\phi({\bf y})}{\partial{\nu}} -\frac{\partial G_0({\bf x}, {\bf y})}{\partial{\nu}_{{\bf y}}} \phi({\bf y})
\right]dS_{{\bf y}} + \sum_{k=1}^{N_{c}}q_k G_0({\bf x}, {\bf y}_k), \quad {\bf x} \in \Omega_1, \\
\label{eqbim_2}
\phi({\bf x}) = 
&\int_\Gamma\left[-G_\kappa({\bf x},
{\bf y})\frac{\partial\phi({\bf y})}{\partial{\nu}} + \frac{\partial G_\kappa({\bf x}, {\bf y})}{\partial{\nu}_{{\bf y}}}
\phi({\bf y})\right]dS_{{\bf y}}, \quad {\bf x} \in \Omega_2,
\end{align}
\end{subequations}
where $G_0({\bf x},{\bf y})$ and $G_\kappa({\bf x}, {\bf y})$ are the 
Coulomb and screened Coulomb potentials,
\begin{equation}
G_0({\bf x}, {\bf y}) = \frac{1}{4\pi|{\bf x}-{\bf y}|},
\quad
\label{eq_potential}
G_\kappa({\bf x}, {\bf y}) = \frac{e^{-\kappa|{\bf x}-{\bf y}|}}{4\pi|{\bf x}-{\bf y}|},
\end{equation}
and ${\bf y}_k \in \Omega_1$ are the location of the atomic centers.

Applying the interface conditions in Eq.~(\ref{eqJump}) 
with the differentiation of electrostatic potential in each domain
yield 
a set of boundary integral equations relating the
surface potential $\phi_1$ (the subscript 1 denotes the inside domain)
and
its normal derivative $\partial\phi_1/\partial{\nu}$ on $\Gamma$, \cite{Juffer:1991, Geng:2013b},
\begin{subequations}
\begin{align}
\label{eqbim_3}
\frac{1}{2}\left(1+\varepsilon\right)\phi_1({\bf x}) & =
\int_\Gamma \left[K_1({\bf x}, {\bf y})\frac{\partial\phi_1({\bf y})}{\partial{\nu}} +
K_2({\bf x}, {\bf y})\phi_1({\bf y})\right]dS_{{\bf y}}+S_{1}({\bf x}),
\quad {\bf x}\in\Gamma, \\
\label{eqbim_4}
\frac{1}{2}\left(1+\frac{1}{\varepsilon}\right)\frac{\partial\phi_1({\bf x})}{\partial{\nu}} & =
\int_\Gamma \left[K_3({\bf x}, {\bf y})\frac{\partial\phi_1({\bf y})}{\partial{\nu}} +
K_4({\bf x}, {\bf y})\phi_1({\bf y})\right]dS_{{\bf y}}
+S_{2}({\bf x}),
\quad {\bf x} \in \Gamma,
\end{align}
\end{subequations}
where $\varepsilon = \varepsilon_2/\varepsilon_1$, 
and
the kernels $K_{1,2,3,4}$ and source terms $S_{1,2}$ are
\begin{subequations}
\begin{align} 
\label{Eq_K1}
K_1({\bf x}, {\bf y}) = 
&\,{G_{0}({\bf x},{\bf y})}-{G_{\kappa}({\bf x},{\bf y})},
\quad
K_2({\bf x}, {\bf y}) = \varepsilon\frac{\partial G_{\kappa}({\bf x},{\bf y})}{\partial\nu_{{\bf y}}}-\frac{\partial
G_{0}({\bf x},{\bf y})}{\partial\nu_{{\bf y}}}, \\
\label{Eq_K2}
K_3({\bf x}, {\bf y}) = 
&\,\frac{\partial G_{0}({\bf x},{\bf y})}{\partial\nu_{{\bf x}}}-\frac{1}{\varepsilon}\frac{\partial G_{\kappa}({\bf x},{\bf y})}{\partial\nu_{{\bf x}}},
\quad
K_4({\bf x}, {\bf y}) =
\frac{\partial^2 G_\kappa({\bf x},{\bf y})}{\partial\nu_{{\bf x}}\partial\nu_{{\bf y}}}-\frac{\partial^2G_{0}({\bf x},{\bf y})}{\partial\nu_{{\bf x}}\partial\nu_{\bf y}},
\end{align}
\end{subequations}
and 
\begin{equation}
S_{1}({\bf x}) = \frac{1}{\veps_1}\sum_{k=1}^{N_{c}}q_kG_{0}({\bf x}, {\bf y}_k),
\quad
S_{2}({\bf x}) = \frac{1}{\veps_1}\sum_{k=1}^{N_{c}}q_k
\frac{\partial G_{0}({\bf x},{\bf y}_k)}{\partial\nu_{{\bf x}}}.
\label{source_terms}
\end{equation}
As given in Eqs.~(\ref{Eq_K1}-\ref{Eq_K2}) and (\ref{source_terms}), the kernels $K_{1,2,3,4}$ and source terms $S_{1,2}$ are linear combinations of $G_0$, $G_k$, 
and their first and second order normal
derivatives~\cite{Juffer:1991,Geng:2013b}. Since the Coulomb potential
is singular, the kernels have the following behavior
\begin{equation*}
  K_1({\bf x}, {\bf y}) = O(1),\;
  K_{2,3,4}({\bf x}, {\bf y}) = O\left(\frac{1}{|{\bf x}-{\bf y}|}\right),\;
\end{equation*}
as $ {\bf y} \to  {\bf x}$.



After the potentials $\phi_1$ and $\partial \phi_1/\partial \nu$ have
been found by solving the boundary integral equation, the  electrostatic
solvation energy can be obtained by
\begin{equation}
E_{\rm sol} = \frac{1}{2}\sum_{k=1}^{N_c}q_k\phi_{\rm reac}({\bf y}_k) =
\frac{1}{2}\displaystyle \sum\limits_{k=1}^{N_c} q_k
\int_\Gamma \left[K_1({\bf y}_k, {\bf y})\frac{\partial\phi_1({\bf y})}
{\partial\nu}+K_2({\bf y}_k, {\bf y})\phi_1({\bf y})\right]dS_{{\bf y}},
\label{solvation_energy}
\end{equation}
where $\phi_{\rm reac}({\bf x}) = \phi_1({\bf x})-S_1({\bf x})$ is the
reaction potential \cite{Juffer:1991,Geng:2013b}. 

\subsection{Galerkin Discretization}

In solving the boundary integral PB equation, 
both the molecular surface and the solution function need to be discretized. 
The molecular surface $\Gamma$ is usually approximated by
a collection of triangles
\begin{equation}
\label{eqbemesh}
\Gamma_{N}=\bigcup\limits^{N}_{i=1} \tau_{i},
\end{equation}
where $N$ is number of elements 
and $\tau_i$ for $i=1,\dots,N$ is a planar triangular boundary element with mid-point $\textbf{x}^c_i$.
This triangulation must be conforming, i.e., the intersection of two
different triangles is either empty, or a common vertex or edge.
Fortunately, surface generators are available, and
our choice for our computations is MSMS~\cite{Sanner:1996}, though
other packages could also be used.
Here, the resolution of the surface can be controlled by the parameter
$d$ that controls the number of vertices per \AA$^2$.
For example, Fig.~\ref{fig_1} (b)
shows the triangulated molecule surface
of the protein barstar, which will bind another protein barnase to form a
biomolecular complex (PDB: 1b2s) \cite{Dong:2003}.

Each triangle $\tau_i$ of $\Gamma_N$ is the parametric image of the reference triangle $\tau$
\begin{equation}
\label{eqreftri}
\tau = \big\{ \eta=(\eta_1, \eta_2) \in \mathbb{R}^2 : 0 \le \eta_1 \le 1 , 0 \le \eta_2 \le \eta_1 \big\}.
\end{equation}
If $\textbf{u}_i$, $\textbf{v}_i$ and $\textbf{w}_i$ are the vertices
then the parameterization is given by
\begin{equation}
\label{eqparameter}
\textbf{x}(\eta) = \textbf{u}_i + \eta_1(\textbf{v}_i-\textbf{u}_i) + \eta_2(\textbf{w}_i-\textbf{u}_i) \in \tau_i \quad \text{for } \eta=(\eta_1,\eta_2)\in\tau.
\end{equation}
The area of the element, the local mesh size, and the global mesh size of the boundary elements $\tau_i$ are given as
$A_i=\frac{1}{2}|(\textbf{v}_i-\textbf{u}_i)\times (\textbf{w}_i-\textbf{u}_i)|$, $h_i=\sqrt{A_i}$, and 
$h=\max\limits_{1\le i \le N}h_i$.

Since a function $f$ defined on $\tau_i$ can be interpreted as a function $g(\eta)$ with respect to the reference element $\tau$,
\begin{equation}
\label{eqf2g}
f(\textbf{x}) = f(\textbf{x}(\eta)) = g(\eta) \quad \text{for } \eta\in\tau, \quad {\bf x}\in\tau_i.
\end{equation}
we can define a finite element space by functions on $\Gamma_N$ whose
pullbacks to the reference triangle are polynomials in $\eta$. The
simplest example is the space of piecewise constant functions, which
are polynomials of order zero on each triangle, which will be denoted
by $S^0_h(\Gamma_N)$. Obviously, the dimension of this space is $N$
and the basis is given by the box functions
\begin{equation}
\label{eqpiecewiseconstant}
\psi^0_i(\bf{x}) =\begin{cases}
1 & \text{if $\bf{x}\in\tau_i$},\\
0 & \text{otherwise},
\end{cases}
\end{equation}
where $i$ is an index of a triangle.

The next step up are piecewise linear
functions. Since there are three independent linear functions on
$\tau$, namely,
\begin{equation}
\label{eqlinearfcn}
\psi^{1}_1(\eta) = 1-\eta_1, \quad \psi^{1}_2(\eta) = \eta_1-\eta_2, \quad \psi^{1}_3(\eta) = \eta_2
\quad \text{for }\eta=(\eta_1, \eta_2)\in\tau.
\end{equation}
the dimension is $3N$. Usually, one works with the space of
continuous linear functions, denoted by $S^1_h(\Gamma_N)$. It is not hard to see that the dimension
of this space is the number of vertices and that the basis is given by
\begin{equation}
\label{eqglinearcfcn}
\psi^c_{i}(\textbf{x}) = \begin{cases}
1 & \text{for $\textbf{x}=\textbf{v}_i$,}\\
0 & \text{for $\textbf{x}=\textbf{v}_j\neq\textbf{v}_i$,}\\
\text{piecewise linear}& \text{elsewhere,}
\end{cases}
\end{equation}
where ${v}_i$ is the $i$-th vertex. 

The approximation powers of piecewise polynomial spaces are well
known, see, e.g., \cite{Rjasanow:2006}. For a function
$w\in H^1_{pw}(\Gamma_N)$ we denote by $w_h^0$ the $L_2$-orthogonal
projection of $w$ into the space of piecewise constant functions, then
\begin{equation}
\label{eqconstanterr}
\|w-w_h^0\|_{L_2(\Gamma_N)} \le
c\left(\sum_{i=1}^{N}h^2_i|w|^2_{H^1(\tau_i)}\right)^{\frac{1}{2}}
  \le c h |w|_{H^1_{pw}(\Gamma_N)},
\end{equation}
where $c$ is the upper bound of the mesh ratio $h_{\max}/h_{\min}$,
$h$ is the maximal diameter of a triangle
and
\begin{equation}
\label{eqHpw}
|w|_{H^1_{pw}(\Gamma_N)} = \Bigg(\sum_{i=1}^{N}|w|^2_{H^1(\tau_i)}\Bigg)^{1/2}.
\end{equation}
Thus, the constant piecewise basis function can give a convergence rate of maximum $O(h)$.

Likewise, the error for the $L_2$-orthogonal projection $w_h^1$ of $w\in
H^2_{pw}(\Gamma_N)$ into $S^1_h(\Gamma_N)$ is
\begin{equation}
\label{lineerr}
\|w-w_h^1\|_{L_2(\Gamma_N)} \le
c\left(\sum_{i=1}^{N}h^2_i|w|^2_{H^2(\tau_i)}\right)^{\frac{1}{2}}
  \le c h^2 |w|_{H^2_{pw}(\Gamma_N)},
\end{equation}

Finite element spaces with higher order polynomials could also be
considered, however their practical value for surfaces with low
regularity and complicated geometries is limited.

The Galerkin discretization is based on a variational formulation of
integral equations (\ref{eqbim_3}) and (\ref{eqbim_4}). That is,
instead of understanding the equations pointwise for ${\bf x} \in \Gamma_N$
the equations are multiplied by test functions $\psi$ and
$\psi^\nu$ and integrated again over $\Gamma_N$. Solving the
variational form amounts to finding $\phi_1$ and
$\partial\phi_1/\partial \nu$ such that 
{\small
\begin{subequations}
\begin{align}
\label{eqbim_5}
\int_{\Gamma_N} \bigg\{ \frac{1}{2}\left(1+\varepsilon\right)\phi_1({\bf x}) & -
\int_{\Gamma_N} \left[K_1({\bf x}, {\bf y})\frac{\partial\phi_1({\bf y})} {\partial{\nu}} +
K_2({\bf x}, {\bf y})\phi_1({\bf y})\right]dS_{{\bf y}} \bigg\} \psi({\bf x}) dS_{\bf{x}}= \int_{\Gamma_N} S_{1}({\bf x}) \psi({\bf x}) dS_{\bf{x}},\\
\label{eqbim_6}
\int_{\Gamma_N} \bigg\{ \frac{1}{2}\left(1+\frac{1}{\varepsilon}\right)\frac{\partial\phi_1({\bf x})}{\partial{\nu}} & -
\int_{\Gamma_N} \left[K_3({\bf x}, {\bf y})\frac{\partial\phi_1({\bf y})}{\partial{\nu}} +
K_4({\bf x}, {\bf y})\phi_1({\bf y})\right]dS_{{\bf y}} \bigg\} \psi^\nu({\bf x}) dS_{\bf{x}}
= \int_{\Gamma_N} S_{2}({\bf x}) \psi^\nu({\bf x}) dS_{\bf{x}},
\end{align}
\end{subequations}
}
holds for all test functions $\psi, \psi^\nu$. In the Galerkin method the solution and
the test functions are formally replaced by functions in the finite element
space. To that end, the unknowns are expanded by basis functions
$\psi_i$ (which
could be either box or hat functions)
\begin{equation}
  \label{eqprojectphi}
 \phi \approx \sum\limits_{i=1}^{N} \phi_{i}\psi_{i}, \quad\mbox{and}\quad
 \frac{\partial \phi}{\partial \nu} \approx \sum\limits_{i=1}^{N} \phi^\nu_{i}\psi_{i}
\end{equation}
and integral equations are tested against the basis functions. This
leads to the linear system $Ax=b$ where $x$ contains the coefficients
in \eqref{eqprojectphi} and
\begin{equation}\label{def:linsys}
A=\left[\begin{array}{cc}A_{11} & A_{12} \\A_{21} & A_{22}\end{array}\right]
\quad \mbox{and} \quad                                               
b =   \left[\begin{array}{c} b_1 \\ b_2 \end{array}\right] .
\end{equation}
The entries of these block matrices are given as
\begin{eqnarray}
\label{eq_linsysEntry}
\nonumber
  A_{11}(i,j) &=& \int_{\Gamma_N}
  \frac{1}{2}\left(1+{\varepsilon}\right) \psi_i({\bf x}) \psi_j({\bf x}) \text{d}S_{\bf x}
+ \int_{\Gamma_N} \int_{\Gamma_N} K_2({\bf x}, {\bf y}) 
   \psi_i({\bf x}) \psi_j({\bf y}) \text{d} S_{\bf y}  \text{d}S_{\bf x}\\\nonumber
  A_{12}(i,j) &=& \int_{\Gamma_N} \int_{\Gamma_N} K_1({\bf x}, {\bf y}) 
   \psi_i({\bf x}) \psi_j({\bf y}) \text{d} S_{\bf y} \text{d}S_{\bf x}\\\nonumber
  A_{21}(i,j) &=& \int_{\Gamma_N} \int_{\Gamma_N} K_4({\bf x}, {\bf y}) 
   \psi_i({\bf x}) \psi_j({\bf y}) \text{d} S_{\bf y}  \text{d}S_{\bf x}\\
  A_{22}(i,j) &=& \int_{\Gamma_N}
  \frac{1}{2}\left(1+\frac{1}{\varepsilon}\right) \psi_i({\bf x}) \psi_j({\bf x}) \text{d}S_{\bf x}
+ \int_{\Gamma_N} \int_{\Gamma_N} K_3({\bf x}, {\bf y}) 
   \psi_i({\bf x}) \psi_j({\bf y}) \text{d} S_{\bf y}  \text{d}S_{\bf x} 
\end{eqnarray}
and the right hand side is
\begin{equation}
  b_1(i) = \int_{\Gamma_N} S_{1}({\bf x}) \psi_i({\bf x}) dS_{\bf{x}}
  \quad \mbox{and} \quad
  b_2(i) = \int_{\Gamma_N} S_{2}({\bf x}) \psi_i({\bf x}) dS_{\bf{x}}
\end{equation}
Since the basis functions vanish on most triangles, the integrations for
the coefficients are only local. For instance, for piecewise constant
elements, the integral $\int_{\Gamma}\dots\psi_i  dS_{\bf{x}}$ reduces
to $\int_{\tau_i}\dots dS_{\bf{x}}$.
Since the coefficients cannot be expressed in analytical form they
have to be calculated by a suitable choice of quadrature rule.
However singularities will appear 
if triangles $\tau_i$ and $\tau_j$ are identical or sharing common edges and vertices.  
To overcome this issue,
we apply the singularity removing transformation of
\cite{Sauter:1996}. This results smooth integrals over a four
dimensional cube. The latter integrals are then approximated by
tensor product Gauss-Legendre quadrature.

After the solution of the linear system has been obtained, 
the electrostatic free solvation energy can be calculated using the
approximations for the surface potentials and its normal derivative
\begin{equation}
\label{eqsolengdis}
E_{sol} = \frac{1}{2}\sum_{n=1}^{N_c}q_n\sum_{i=1}^{N}
\int_{\tau_{i}}\Big[K_1(\textbf{x}_n,\textbf{x})  \phi_{1i}^{\nu} + K_2(\textbf{x}_n,\textbf{x})\phi_{1i} \Big] dS_{\textbf{x}}.
\end{equation}


Since the matrix $A$ is a dense and non-symmetric our choice of solver
is the GMRES method.
In each step of GMRES iteration, a matrix-vector product is calculated and a direct summation for this requires $O(N^2)$ complexity.
Below we will introduce the $O(N)$ Cartesian Fast Multipole Method (FMM) to accelerate the matrix-vector product. Calculating the electrostatic solvation free energy $E_{sol}$ in Eq.~(\ref{eqsolengdis}) is $O(N_cN)$ and we use a Cartesian treecode to reduce the cost to $O(N_c\log(N))$. Both FMM and treecode algorithm are described the next for comparison and for the reason that both are used to accelerate the N-body particle-particle interactions.


\subsection{Cartesian fast multipole method (FMM)}
In this section, 
we introduce the Cartesian FMM to evaluate matrix vector products with
the matrix in (\ref{def:linsys}) efficiently. This is a
kernel-independent version of the FMM, where instead of the multipole
expansion, truncated Taylor series are used to approximate  Coulomb ($\kappa=0$) the screened Coulomb ($\kappa \ne 0$) potential.

This considerably simplifies the translation operators for the kernels $K_{1-4}$ because
they involve different values of $\kappa$. It was shown in
\cite{Tausch:2003} how the derivatives of the Coulomb kernel can be
computed by simple recurrence formulas. 
Furthermore, the moment-to-moment (MtM) and local-to-local (LtL) translation
are easily derived using the binomial formula. 

When a refined mesh is required for a larger $N$, 
increasing the expansion order is essential to control the accuracy.
The FMM error analysis implies that the truncated Taylor expansion error 
has the same magnitude as the discretization error if the expansion
order is adjusted to the level according to the formula
\begin{equation}\label{variable:order}
  p_l = p_L + L - l
\end{equation}
where $l = 0,1,\cdots,L$ with $l=0$ the coarsest level and $l=L$ the finest level. That
is, the finest level uses a low-order expansion $p_L$, and the order
is incremented in each coarser level, see \cite{Tausch:2004}.

Note that the multipole series is more efficient as it contains $(p+1)^2$
terms, while the Taylor series has $p(p+1)(p+2)/6$ terms. This
difference becomes significant with larger values of $p$. However,
with the variable order scheme the advantage of the multipole series
is much less because most translation operators are in the fine levels
where the number of terms in both series are comparable.

The matrix vector product can be considered as generalized $N$-body
problem of the form
\begin{equation}
\label{eqgeneralKernel}
V_i = \int_{\Gamma_N} \int_{\Gamma_N} \psi_i({\bf x})\frac{\partial^l}{\partial\nu_{\bf x}^l} \frac{\partial^k}{\partial\nu_{\bf y}^k} G({\bf x},{\bf y}) f_h({\bf y}) dS_{\bf y} dS_{\bf x}.
\end{equation}
where $k,l \in \{0,1\}$ and $f_h$ is a linear combination of the basis
functions $\psi_j$.



Next we show how the Cartesian FMM is used under the framework of boundary element method. 

\begin{figure}[htb]
\begin{center}
\includegraphics[width=3.3in]{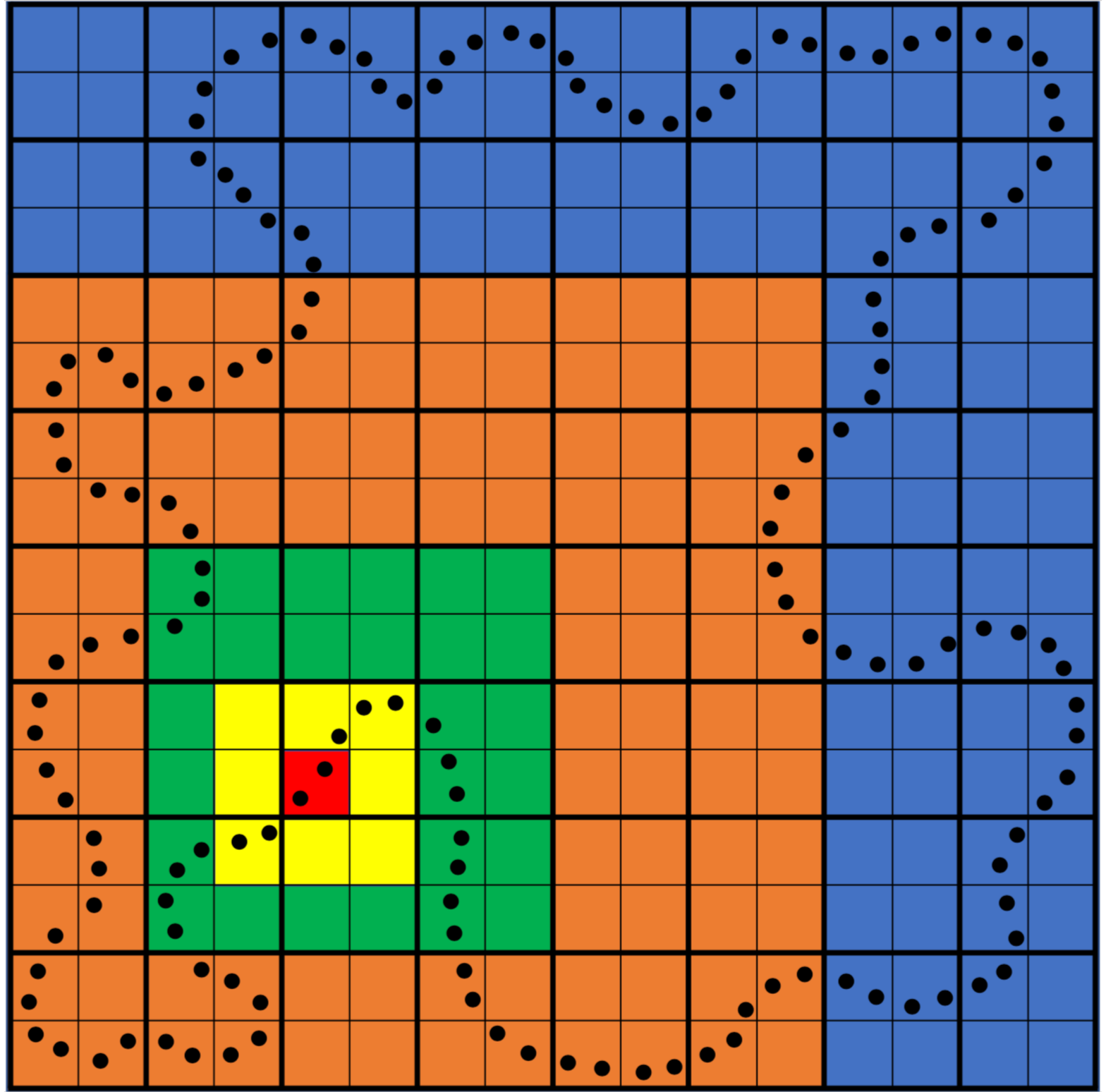}
\includegraphics[width=3.3in]{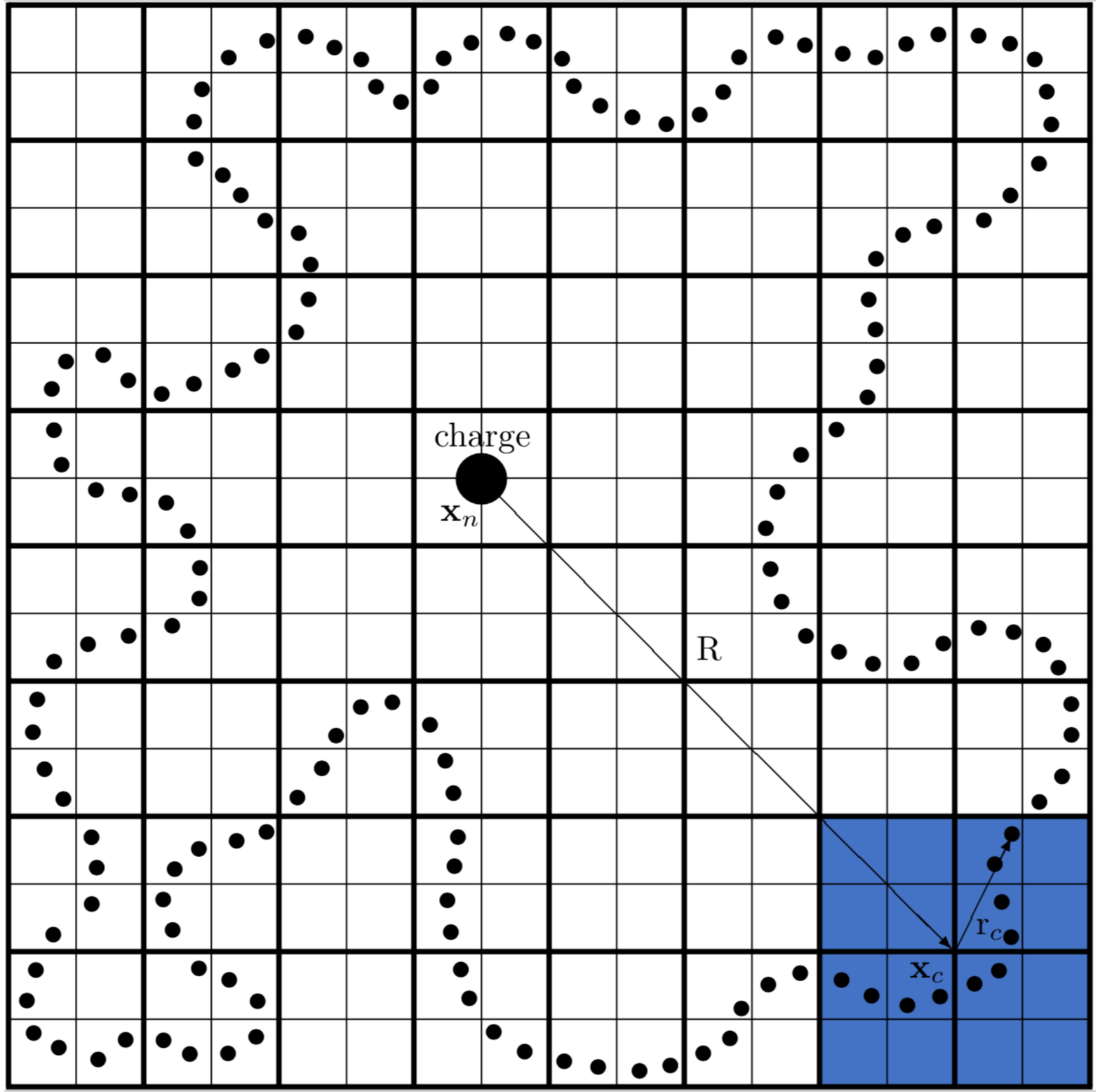}
\caption{FMM vs Treecode structure. 
Left: FMM cluster-cluster interaction list; 
Right: treecode particle-cluster interaction ($R$ is the distance from the charge to the blue cluster's center;
$r_c$ is the radius of the blue cluster $c$ which is the farthest particle inside $c$ to the center of $c$.}
\label{fig_2}
\end{center}
\end{figure}

Figure~\ref{fig_2}a is a 2-D illustration of a discretized molecular surface $\Gamma_N$ 
embedded in a hierarchy of cubes (squares in the image). 
Each black solid dot represents a boundary triangle  $\tau_{i}$ for $i=1,\cdots,N$. 

A cluster $c$ in any level $l$ is defined as the union of triangles
whose centroid are located in a cube of that level. $C_l$ is the set
of all clusters in level $l$.


The level-0 cube is the smallest axiparallel cube that contains
$\Gamma_N$ and 
thus $C_0 = \Gamma_N$. The refinement of coarser cubes into finer
cubes stops when clusters in the finest level contain at most a
predetermined (small) number of triangles. For a cluster $c$ we denote
by $B_c$ the smallest axiparallel rectangular box that contains $c$
and write ${\bf x}_c$ for its center and $\rho_c$ for the half-length
of its longest diagonal. Note that $B_c$ can be
considerably smaller than its cube, this is why we call this process
the shrink scheme. For two clusters $c$ and $c'$ in the same level we
denote by
\begin{equation}
  \eta(c,c') = \frac{ \rho_c + \rho_{c'}}{|{\bf x}_c - {\bf x}_{c'}|}\label{def:eta}
\end{equation}
the separation ratio of the two clusters. This number determines the
convergence rate of the Taylor series
expansion, see~\cite{Tausch:2004}. Two clusters in the same level are neighbors if
their separation ratio is larger than a predetermined constant.  
$\mathcal{N}(c)$ denotes the set of its neighbors for a given cluster $c$.
The set of nonempty children of $c$ that are generated in the refinement
process is denoted by $\mathcal{K}(c)$.
Finally, we use $\mathcal{I}(c)$ to denote the interaction list for a cluster $c$,
which are clusters at the same level such that 
for any $c'\in \mathcal{I}(c)$, 
the parent of $c'$ is a neighbor of the parent of $c$, but $c'$ itself
is not a neighbor of $c$.

Under the FMM framework, 
the evaluation of Eq.~(\ref{eqgeneralKernel}) consists of the near field direct summation 
and the Taylor expansion approximation for well-separated far field.
The near field direct summation happens in between neighboring panels
in the finest level.  The far field summation is done by multipole or
Taylor expansions between interaction lists in all levels. This
process is described in many papers, so we do not give details about
the derivation. 


To emphasize the differences of the cartesian FMM, we 
consider a cluster-cluster interaction between two clusters $c$ and
$c' \in \mathcal{I}(c)$. Let $u$ be the potential due to sources in
$c'$ which is evaluated in $c$, then  
%
%
by Taylor expansion of the kernel with center ${\bf x} = {\bf x}_c$ and ${\bf y} =
{\bf x}_{c'}$ one finds easily that
\begin{equation}
\label{eqapproximation}
u_{c,c'}({\bf x}) = \int_{c'} \frac{\partial^l}{\partial\nu_{\bf x}^l} \frac{\partial^k}{\partial\nu_{\bf y}^k} G({\bf x}, {\bf y})f_h({\bf y}) dS_{\bf y} \approx \sum_{|\alpha| \leq p}\lambda ^{\alpha}_{c}  \frac{\partial^l}{\partial\nu_{\bf x}^l} ({\bf x}-{\bf x}_{c})^{\alpha} .
\end{equation}
where 
and $\alpha = (\alpha_1, \alpha_2, \alpha_3) \in \mathbb{N}^3$
is a multi-index. 
The expansion coefficients are given by 
\begin{equation}
\label{eqcoeff}
\lambda^{\alpha}_{c} = \sum\limits^{p-|\alpha|}_{|\beta|=0}\frac{D^{\alpha+\beta} G({\bf x}_{c}, {\bf x}_{c'})}{\alpha!\beta!}(-1)^{\beta} m^{\beta}_{c'}(f_h), \quad |\alpha| \leq p.
\end{equation}
where $\al!=\al_1!\al_2!\al_3!$ and $m^{\beta}_{c'}(f)$ is the
moment of $f_h$, given by
\begin{equation}
\label{eqmoment}
m^{\beta}_{c'}(f) = \int_{S_{c'}} \frac{\partial^k}{\partial\nu_{\bf x}^k} ({\bf x} - {\bf x}_{c'})^{\beta}f_h({\bf x})dS_{\bf x}, \quad |\beta| \leq p.
\end{equation}
Equation~(\ref{eqcoeff}) translates the moment of $c'$ to the local expansion coefficients of the cluster $c$,
and it is therefore called MtL translation. Since $f_h$ is a linear
combination of basis functions, we obtain from linearity that
\begin{equation}\label{eqmomentL}
m^{\beta}_{c'}(f) = \sum_{i\in c'} m^{\beta}_{c'}(\psi_i) f_i
\end{equation}
where $f_i$ are the coefficients of $f_h$ with respect to the
$\psi_i$-basis and the summation is taken over basis functions that
whose support overlaps with $c$.
Since we consider a Galerkin discretization we have to integrate the
function $u_{c,c'}({\bf x})$ against the test functions, to obtain the
contribution $g_i$ of the two clusters to the matrix vector product. Thus we
get from (\ref{eqapproximation})
\begin{equation}\label{LtP}
g_i = \int_{c} \psi_i({\bf x}) u_{c,c'}({\bf x}) dS_{\bf x} =
\sum_{|\alpha| \leq p}\lambda ^{\alpha}_{c}  \int_{c}
\frac{\partial^l}{\partial\nu_{\bf x}^l} ({\bf x}-{\bf
  x}_{c})^{\alpha}  \psi_i({\bf x}) dS_{\bf x} =
\sum_{|\alpha| \leq p}\lambda^{\alpha}_{c} m_c^{\alpha}(\psi_i)
\end{equation}
This operation converts expansion coefficients to potentials and is
denoted as LtP translation.

To move moments and local expansion coefficients between levels, we also
need the moment-to-moment (MtM) and local-to-local (LtL)
translations. They can be derived easily from the multivariate
binomial formula. We obtain
%
\begin{equation}\label{eqMtM}
m^{\alpha}_{c}(f)
= \sum_{c'\in\mathcal{K}(c)}\sum_{\beta \leq \alpha} \binom {\al}{\ba} ({\bf x}_{c'} - {\bf x}_{c})^{\al-\ba} m^{\beta}_{c'}(f),
\end{equation}
and 
\begin{equation}\label{eqLtL}
\lambda ^{\beta}_{c'} = \sum_{\substack{\al\le\ba \\ |\al|\le p}} \binom {\al}{\ba} ( {\bf x}_{c'} - {\bf x}_{c})^{\al-\ba} \lambda ^{\alpha}_{c},  \quad |\ba|\le p.
\end{equation}
where $c' \in \mathcal{K}(c)$.

We see that moments and expansion coefficients are computed by
recurrence from the previous level. In the finest level the moments of
the basisfunctions $m_c(\psi_i)$ can be either computed by numerical
quadrature, or even analytically, because we consider flat panels and
polynomial ansatz functions. We skip the details, as these formulas are
straightforward application of the binomial formula. 

In summary, the Cartesian FMM under the framework of boundary element method is described as the following.

\begin{enumerate}
\item \emph{Nearfield Calculation.}\\
for $c \in C_L$\\
\hspace*{1.5em} for $c' \in \mathcal{N}(c)$ \\
\hspace*{3em} multiply matrix block of $c$ and $c'$ directly.

\item \emph{Moment Calculation.} \\
for $c \in C_L$\\
\hspace*{1.5em} Compute the moments $m^{\beta}_{c'}(f)$ in (\ref{eqmomentL}).

\item \emph{Upward Pass.} \\
for $l=L-1,\dots,l_{min}$\\
\hspace*{1.5em}for $c \in C_l$\\
\hspace*{3em}for $c' \in \mathcal{K}(c)$\\
\hspace*{4.5em} Compute the MtM translation (\ref{eqMtM})

\item \emph{Interaction Phase.} \\
for $l=L,\dots,l_{min}$\\
\hspace*{1.5em} for $\nu \in C_l$\\
\hspace*{3em} for $\nu' \in \mathcal{N}(c)$\\
\hspace*{4.5em} Compute the MtL translation (\ref{eqcoeff})

\item \emph{Downward Pass.} \\
for $l=l_{min},\dots,L-1$\\
\hspace*{1.5em} for $c \in C_l$\\
\hspace*{3em} for $c' \in \mathcal{K}(c)$\\
\hspace*{4.5em} Compute the LtL translation (\ref{eqLtL})

\item \emph{Evaluation Phase.}\\
for $c \in C_L$\\
\hspace*{1.5em} Compute the LtP translation (\ref{LtP})
\end{enumerate}

In this algorithm $l_{min}$ is the coarsest level that contains
clusters with non-empty interaction lists.

Since the finest level contain a fixed number of triangles the number
of levels grows logarithmically with $N$ as the mesh is refined. 
With a geometric series argument, one can show that the total number
of interaction lists in all levels is $O(N)$. If the translations in
all levels are computed with the same order $p$ then the complexity of
all translations is  $O(N p^3)$. If the variable order method is used
where the order is given by (\ref{variable:order}), then the
complexity reduces to $O(N)$. More details can be found in ~\cite{Tausch:2004}.

\subsection{Cartesian treecode}
The Cartesian treecode can be considered as a fast multipole method
without the downward pass.  
The computational cost of treecode is order of $O(N\log N)$ as opposed to the $O(N)$ FMM.  
However, the constants in this complexity estimate are smaller, and we
found it to be useful for the computation of the free solvation energy
(\ref{eqsolengdis}), where
the source and evaluation points are different and zero or limited near field
calculations are required.
The direct computation of the solvation energy 
as interactions between $N$ boundary elements and $N_c$ atomic centers
has $O(N_cN)$ complexity. 
This is shown in Fig.~\ref{fig_2}b, 
in which a charge located at ${\bf x}_n$ will interact with induced charges ($\phi_1$ or $\frac{\partial \phi_1}{\partial \nu}$)
located at the center of each panel. 
These interactions consist of near field particle-particle interaction by direction summation 
and far field particle-cluster interaction controlled by maximum acceptance criterion (MAC) as specified below. 
For simplicity, we write the involved calculations as 
\begin{equation}
\label{eqsolengform}
E = \sum_{n=1}^{N_c} q_n V_n
 = \sum_{n=1}^{N_c} q_n\sum_{l=1}^{N}
\int_{\tau_{l}} \frac{\partial^k}{\partial\nu_\textbf{x}^k} G(\textbf{x}_n,\textbf{x})f(\textbf{x})dS_\textbf{x},
\end{equation}
where $G$ is Coulomb or screened Coulomb potential kernel, $k \in
\{0,1\}$, 
$q_n$'s are partial charges,
and $f$ is either $\phi_1$ or $\partial\phi_1/\partial\nu$.



\begin{figure}
\begin{subfigure}{.5\textwidth}
\centering
\includegraphics[width=\linewidth]{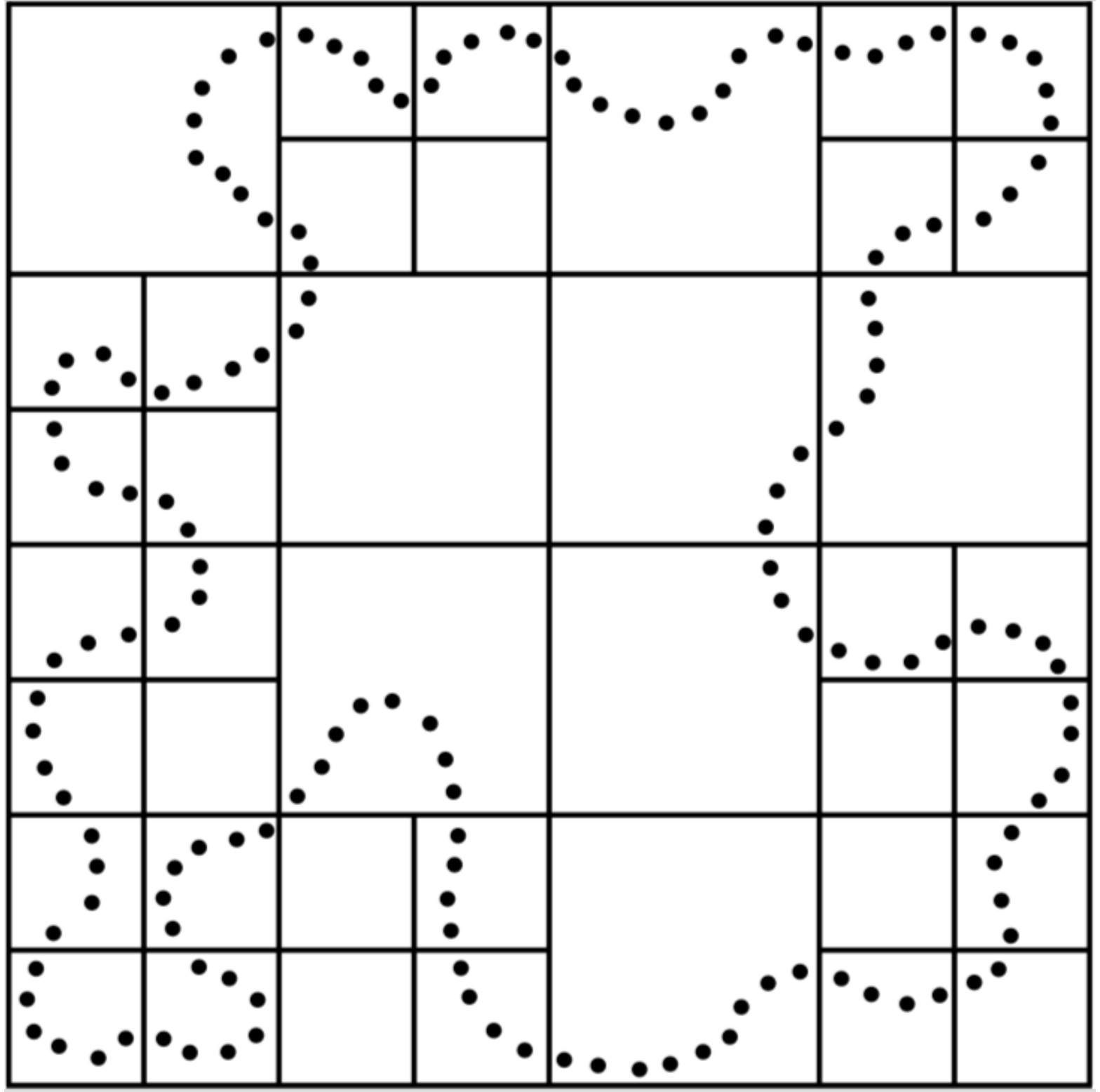}
\caption{}
\label{fig_3_1}
\end{subfigure}
\begin{subfigure}{.5\textwidth}
\centering
\includegraphics[width=\linewidth]{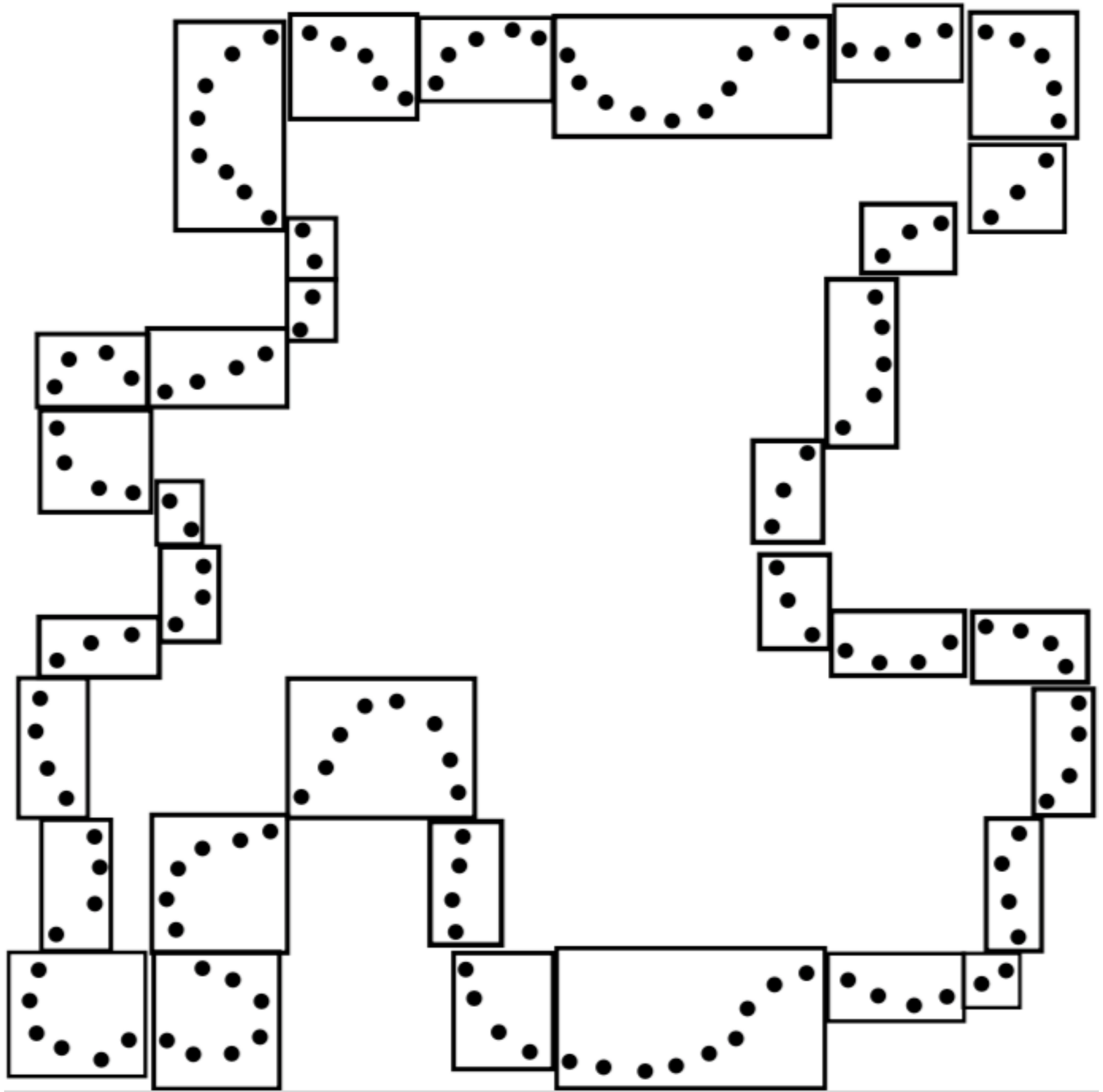}
\caption{}
\label{fig_3_2}
\end{subfigure}
\caption{The hierarchical tree structure for the treecode: Left: the form of the tree by recursively dividing a parent cluster into eight/four children clusters using $N_0=8$; Right:  The adaptive shrink scheme improves the treecode efficiency.}
\label{fig_3}
\end{figure}


In our implementation, we use the same clustering scheme of
$\Gamma_N$ as in the FMM. Instead of using the interaction lists to calculate the far field
interaction, the treecode use the following multipole acceptance criterion (MAC) to determine if the particle and the cluster are well separated or thus a far-field particle-cluster interaction will be considered.  
This is similar to the separation ratio in the FMM. The MAC is given as 
\begin{equation}
\label{eqMAC}
\frac{r_c}{R} \le \theta,
\end{equation}
where $r_c=\text{max}_{\textbf{x}_j\in c}|\textbf{x}_j-\textbf{x}_c|$ is the cluster radius,
$R=|\textbf{x}_n-\textbf{x}_c|$ is the particle-cluster distance,
and $\theta<1$ is a user-specified parameter.
If the criterion is not satisfied,
the program checks the children of the cluster recursively
until either the MAC is satisfied
or the leaves (the finest level cluster) are reached
at which direct summation is applied.
Overall, the treecode evaluates the potentials~(\ref{eqsolengform}) 
as a combination of particle-cluster interactions and direct summations.
Thus, when $\textbf{x}_n$ and $c$ are well-separated,
the potential can be evaluated as
\begin{equation}
\label{eqsolengformTree}
\int_{c} \frac{\partial^{k}}{\partial\nu_{\bf x}^{k}} G({\bf x}_n,{\bf x}) f({\bf x})dS_{\bf x}
\approx 
\sum\limits^{p}_{|\beta|=0} D^{\beta}G({\bf x}_{n}, {\bf x}_{c}) (-1)^{\beta} {m'}^{\beta}_{c}(f),
\end{equation}
where the moment ${m'}^{\beta}_{c}(f)$ is calculated by the same operator-MtM in FMM.

The treecode method therefore can be concluded as

\begin{enumerate}
\item \emph{Moment Calculation.} \\
for $c \in C_L$\\
\hspace*{1.5em} Compute the moments $m^{\beta}_{c'}(f)$ in (\ref{eqmomentL}).

\item \emph{Upward Pass.} \\
for $l=L-1,\dots,l_{min}$\\
\hspace*{1.5em}for $c \in C_l$\\
\hspace*{3em}for $c' \in \mathcal{K}(c)$\\
\hspace*{4.5em} Compute the MtM translation (\ref{eqMtM})

\item \emph{Interaction Phase}\\
  for $n=1,...,N_c$\\
\hspace*{1.5em}$E_n = 0$\\
\hspace*{1.5em}for $c \in C_0$\\
\hspace*{3em} addCluster(c,$\textbf{x}_n$,$E_n$)    
\end{enumerate}
where addCluster(c,$\textbf{x}_n$,$E_n$) as shown below  is a routine that
recurses from the coarse clusters to the finer clusters until the
separation is sufficient to use the Taylor series approximation\\
\begin{itemize}
  \item[]if $\textbf{x}_n$ and $\textbf{x}_c$ satisfy the MAC for $c$\\
	$~~~~~~~~~~~~~~E_n\mathrel{+}=\sum\limits^{p}_{|\beta|=0} D^{\beta}\Phi({\bf x}_{n}, {\bf x}_{c}) (-1)^{\beta} {m'}^{\beta}_{c}(f)$
	\item[] else if $\mathcal{K}(c) \ne \emptyset$\\
          $~~~~~~$for $c'\in \mathcal{K}(c)$\\
          $~~~~~~~~~~~~~~$ addCluster($c'$,$\textbf{x}_n$,$E_n$)
	\item[] else\\ 
	$~~~~~~~~~~~~~~E_n\mathrel{+}=\displaystyle\int_{c} \frac{\partial^{\ka}}{\partial\nu_{\bf x}^{\ka}} \Phi({\bf x}_n,{\bf x})f(x)dS_{\bf x}$
\end{itemize}

Note that steps $1$ and $2$ are analogous to the steps the in FMM,
hence the addition of the treecode to the FMM code requires little
extra work.

\subsection{Preconditioning}
The results in our previous work \cite{Geng:2013b, Geng:2013a} shows
that the PB boundary integral formulation in Eqs.~(\ref{eqbim_3}) and (\ref{eqbim_4})
is well-conditioned thus will only require a small number of GMRES iteration if
the triangulation quality is satisfied (e.g. nearly quasi-uniform). 
However, due to the complexity of the molecular surface, 
the triangulation unavoidably has a few triangles with defects (e.g. narrow triangles and tiny triangles)
which deteriorate the condition number of the linear algebraic matrix, resulting in      
increased GMRES iteration number required to reach the desired convergence accuracy. 
 
Recently, we designed a block-diagonal preconditioning scheme 
to improve the matrix condition for the treecode-accelerated boundary integral (TABI) Poisson-Boltzmann solver \cite{Chen:2018a}.
The essential idea for this preconditioning scheme is 
to use the short range interactions within the leaves of the tree to form the preconditioning matrix $M$. 
This preconditioning matrix $M$ can be permuted into a block diagonal form thus $Mx=y$
can solved by the efficient and accurate direct methods. 
In the current study of FAGBI solver,  
the same conditioning issue rises and it can be resolved 
by a similar but FMM structure adjusted and controlled preconditioning scheme. 

The key idea is to find an approximating matrix $M$ of $A$ such that
$M$ is similar to $A$ and  the linear system $My=z$ is easy to solve.
To this end, our choice for $M$ is the matrix 
involving only direct sum interactions in cubes/clusters at a designated level 
(an optimal choice considering both cost and efficiency) 
as opposed to $A$, 
which involves all interactions.

The definition of $M$ will be essentially similar to $A$ in
(\ref{def:linsys}) except that the entries of $M$ are zero 
if $\tau_i$ and $\tau_j$ are not on the same cube at a designated level of the tree, i.e. 
\begin{equation}
M_{mn}(i,j) = \left\{\begin{array}{l}
A_{mn}(i,j) \text{~~~if~$\tau_i, \tau_j$ are on the same cube at a designated level of the tree}  \\
0 \text{~~~~~~~~~~~~~otherwise.}\end{array}\right. 
\end{equation}

\begin{figure}[htb]
	\begin{center}
		\includegraphics[width=2.18in]{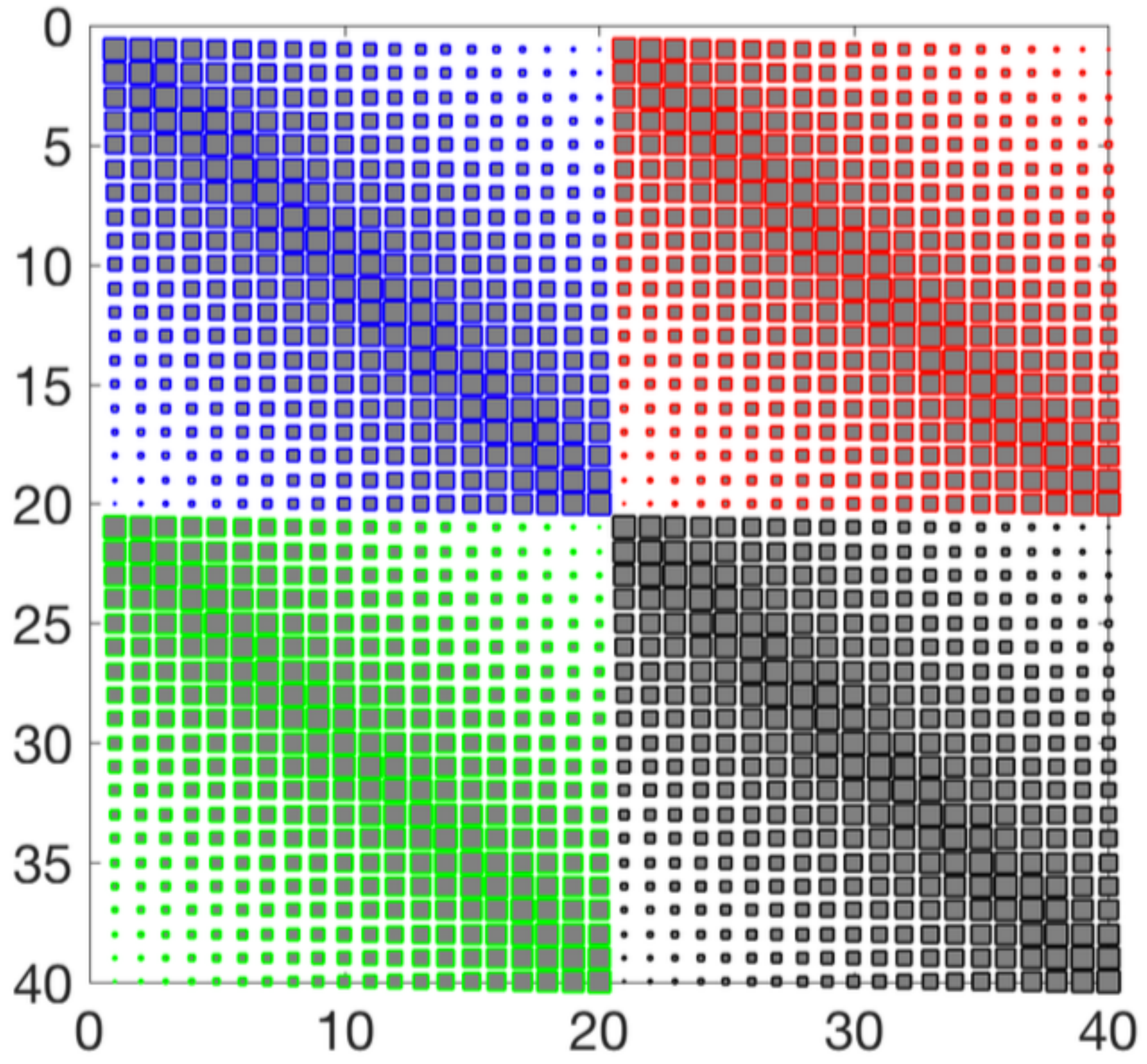}
		\includegraphics[width=2.18in]{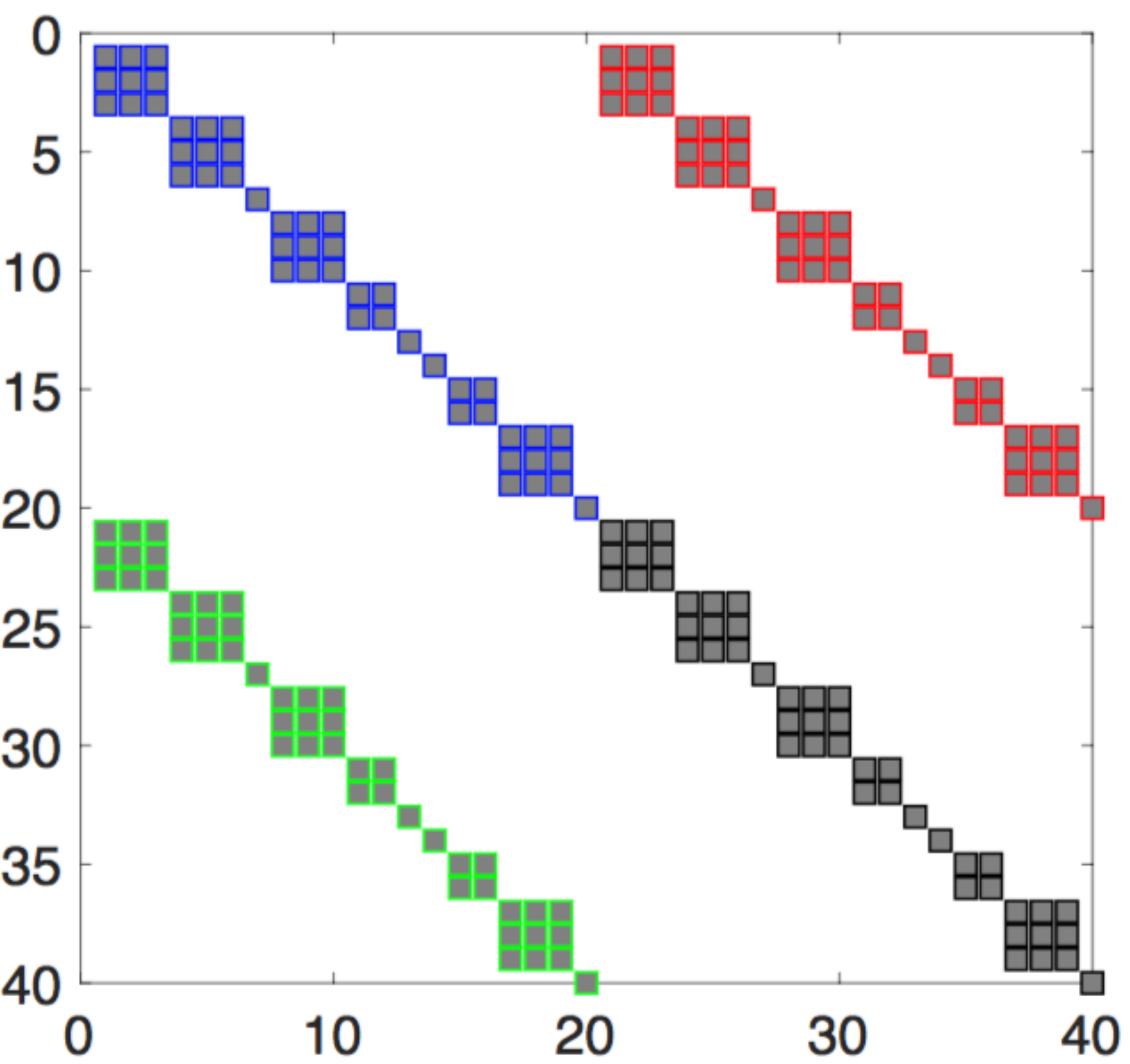}
		\includegraphics[width=2.18in]{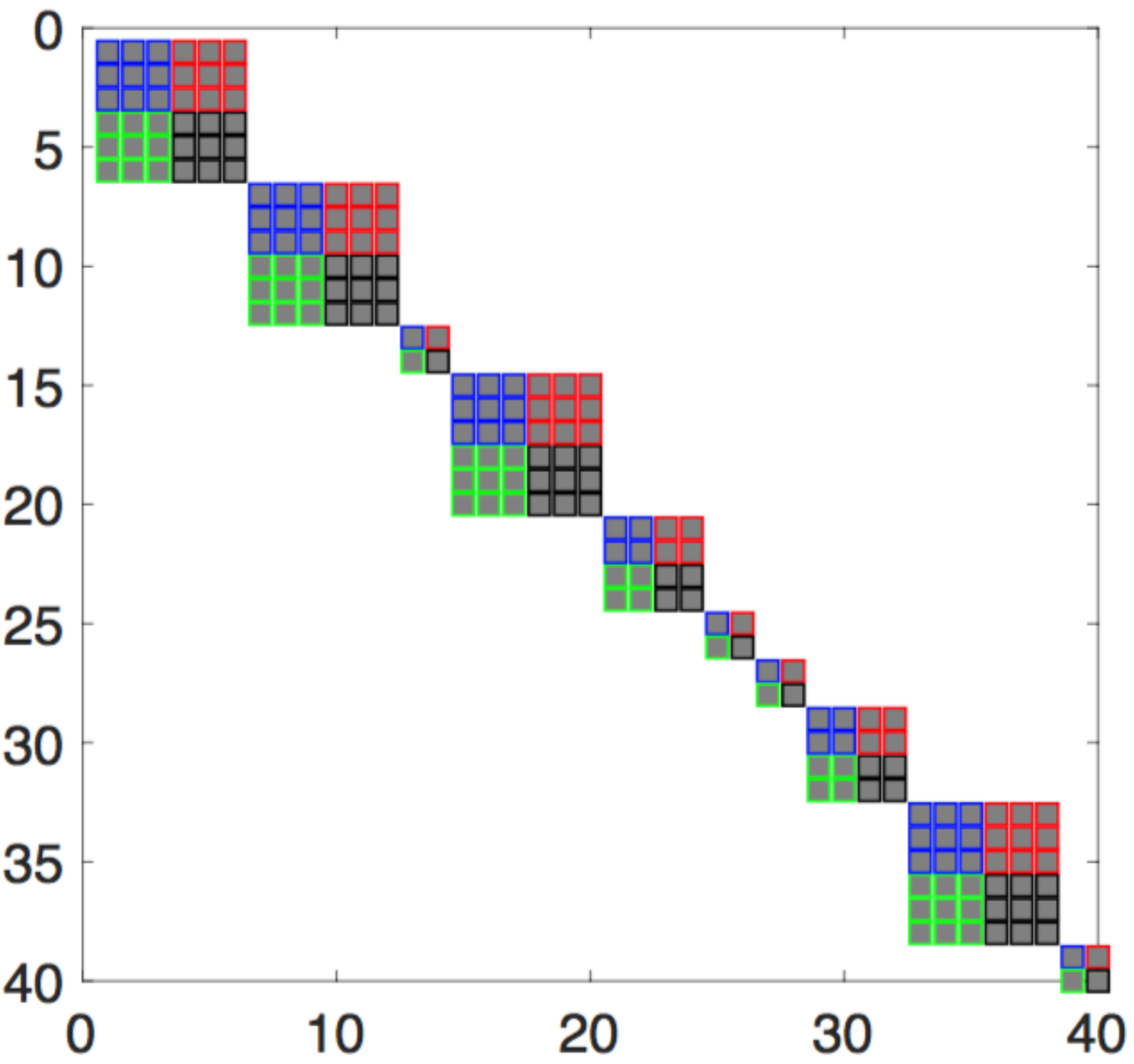}
		\hskip 0.8in (a) \hskip 2.0in (b) \hskip 2.0in (c)
		\caption{A schematic illustration of the boundary element dense matrix $A$ and its preconditioning matrix $M$:  (a) matrix $A$ for the case of $N=20$ elements (the size of the matrix entry shows the strength of the interaction; the four different color-coded region relates to $K_{1-4}$ in Eqs.~(\ref{eqbim_3})-(\ref{eqbim_4})); (b) the ``block diagonal block" preconditioning matrix $M$ (assuming the cube at the designated level contains at most 3 panels; (c)  the ``block diagonal"  preconditioning matrix $M$, which is a permuted matrix from $M$ in (b) after switching the order of the unknowns. 
		}
		\label{fig_matrices}
	\end{center}
\end{figure}

Here we use Fig.~\ref{fig_matrices} to illustrate 
how we design our preconditioning scheme and its advantage.
Figure~\ref{fig_matrices}(a) is the illustration of the dense boundary element matrix $A$
for the discretized system~(\ref{def:linsys}) with 20 boundary elements. 
The four different colors represent the four kernels $K_{1-4}$ related entries of the linear algebraic matrix $A$ in Eq.~(\ref{eq_linsysEntry}). 
Note the unknowns are ordered by the potentials $\phi_1$ on all elements, 
followed by the normal derivative of the potential $\frac{\partial \phi_1}{\partial \nu}$. 
The size of the matrix entry in Fig.~\ref{fig_matrices} indicates the magnitude of the interaction 
between a target element and a source element, which decays from the main diagonal to its two wings. 
By only including the interactions between elements on the same cube at a designated level, 
we obtain our designed preconditioning matrix $M$ as illustrated in 
Fig.~\ref{fig_matrices}(b). This preconditioning matrix $M$ has four blocks, and each block is 
a diagonal block matrix. 
Following the procedure detailed in \cite{Chen:2018a}, 
by rearranging the order of the unknowns, a block diagonal matrix $M$ 
is achieved as illustrated in Fig.~\ref{fig_matrices}(c). 
Since $M=\text{diag}\{M_1, M_2, \cdots, M_{N_l}\}$ as shown in  Fig.~\ref{fig_matrices}(c) is a block diagonal matrix 
such that $My=z$ can be solved using direct method e.g. LU factorization by solving each individual $M_iy_i=z_i$. Here each $M_i$ is a square nonsingular matrix, which represents the interaction between particles/elements on the $i$th cube of the tree at a designated level.  As shown in \cite{Chen:2018a}, the total cost of solving $My=z$ is essentially $O(N)$ thus is very efficient. Results for the preconditioning performance will be shown in the next section. 

\section{Results}
Our numerical results are mostly produced 
on a desktop with an i5 7500 CPU and 16G Memory, 
using GNU Fortran 7.2.0 compiler with compiling option ``-O2''. 
A few results for the long elapsed direct summation are obtained from the SMU high performance computing cluster, ManeFrame II (M2), with Intel Xeon Phi 7230 Processors, using openmpi/3.1.3 compiler with compiling option ``-O2''. 
Note these direct sum results are needed for the evaluation of accuracy only 
and low resolutions results are checked on different machines to ensure the consistency of the accuracy.
All protein structures are obtained from Protein Data Bank (\url{https://www.wwpdb.org)} and partial charges are assigned by CHARMM22 force field \cite{Brooks:1983} using PDB2PQR software \cite{Dolinsky:2007}.

The physical quantity we computed in this manuscript is the electrostatic free energy of solvation with the unit kcal/mol. The electrostatic potential $\phi$ or $\phi_1$ governed in Eq.~(\ref{eqNPBE}) or Eqs.~(\ref{eqbim_3}-\ref{eqbim_4}) uses the unit of $e_c/(4\pi$\AA), where $e_c$ is the elementary charge. By doing this, we can directly use the partial charge obtained from PDB2PQR \cite{Dolinsky:2007} for solving the PB equation. 
After obtaining the potential, we can convert the unit $e_c/(4\pi$\AA) to kcal/mol/$e_c$ by multiplying the constant $4\pi332.0716$ at room temperature T=300K. From potential to energy, only a multiplication of $e_c$ is needed. 

We solved the PB equation first on the Kirkwood sphere \cite{Kirkwood:1934}, where the analytic solution is available to validate the accuracy and efficiency of FAGBI solver, then on a typical protein 1a63 to demonstrate the overall performance, and finally on a series of proteins to emphasize the preconditioning scheme and the broad usage of the FAGBI solver. 

\subsection{The Kirkwood sphere}
Our first test case is the Kirkwood sphere of radius 50\AA~with an atomic charge $q=50e_c$ at the center of the sphere.
The dielectric constant is $\varepsilon_1=1$ inside the sphere and $\varepsilon_2=40$ outside the sphere. 
We provide three parts of this test case on the Kirkwood sphere, 
which show first the discretization error, 
then the impact of the quadrature orders toward the convergence of accuracy, 
and finally the comparison between using the Cartesian FMM and using direct sum 
in terms of error, CPU time, and memory usage. \\ 

\subsubsection{Overall discretization error}
We first solve the boundary integral PB equation on the Kirkwood sphere 
using the direct summation for matrix-vector product instead of using the FMM acceleration. 
The linear algebraic system is solved using GMRES iterative solver with $L_2$ relative tolerance $\tau = 10^{-6}$. 
The Galerkin method is applied to form the matrix combined with a single point Gauss quadrature. 
Cubature methods \cite{Sauter:1996} are applied for treating the singularities 
arising from Galerkin discretization of boundary integral equations.

\begin{table}[htp!]
\caption{\small Discretization error from solving the PB equation on a Kirkwood sphere with a centered charge.  
Results include electrostatic solvation free energy $E^{ds}_{sol}$ with error $e^{ds}_{sol}$ and convergence rate $r^{ds}_{sol}$, and discretization error in surface potential $e^{ds}_{\phi}$, normal derivative $e^{ds}_{\partial_n \phi}$ with their convergence rates $r^{ds}_{\phi}$ and $r^{ds}_{\partial_n \phi}$.}
{
\small
\begin{center}
\begin{tabular}{lccccccccc}
\hline
\rule{0pt}{12pt}
$N^1$ &$h$& $E^{ds}_{sol}$ (kcal/mol) & $e^{ds}_{sol}$ (\%) & $r^{ds}_{sol}$ & $e^{ds}_{\phi}$ (\%) & $r^{ds}_{\phi}$ & $e^{ds}_{\partial_n \phi}$ (\%) & $r^{ds}_{\partial_n \phi}$ & $\text{Iters}^2$ \\\hline
320      &9.90&-8413.28&1.692&3.6&17.925&4.1&0.652&1.9&3\\
1280     &4.95&-8328.18&0.663&2.6& 4.419&4.1&0.212&3.1&3\\
5120     &2.48&-8293.42&0.243&2.7& 1.112&4.0&0.092&2.3&3\\
20,480   &1.24&-8280.70&0.089&2.7& 0.285&3.9&0.046&2.0&3\\
81,920   &0.62&-8276.33&0.036&2.5& 0.077&3.7&0.023&2.0&3\\
327,680  &0.31&-8274.64&0.016&2.3& 0.022&3.5&0.011&2.0&3\\
1,310,720&0.15&-8273.91&0.007&2.2& 0.007&3.2&0.006&2.0&4\\
$\infty^3$ &&-8273.31&&&&&&&\\
\hline
\end{tabular}
\end{center}
${}^1$ $N$ is number of triangles in triangulation; $h$ is the average of largest edge length of all triangles; $h\approx O(N^{-2})$\\
${}^2$ Number of GMRES iterations.\\
${}^3$ This row displays the exact electrostatic solvation energy $E^{ex}_{sol}$,
which is known analytically.
}
\label{tb_discretization}
\end{table}%

Table~\ref{tb_discretization} shows the total discretization errors, 
which is related to triangulation, quadrature, and basis function. 
In this table, Column 1 is the number of triangles $N$ for the sphere with the refinement of the mesh and 
Column 2 is the average of largest edge length of all triangles $h$. Note we have $h \approx O(N^{-2})$, which can be seen from the comparison of values in the two columns. Since using $N$ is more convenient  to specify mesh refinement in our numerical simulation, we use it to quantify the mesh refinement for the rest of the paper.

Columns 3-4 show that the electrostatic solvation energy $E^{ds}_{sol}$ 
and its error $e^{ds}_{sol}$ compared with the true value in the last row of the table. 
The convergence rate $r^{ds}_{sol}$ defined as the ratio of the error is shown in column 5 with an $O(N^{-1/2})$ pattern.  
The relative $L_\infty$ errors of surface potential $\phi$, $e^{ds}_{\phi}$ 
and normal derivative $\partial_n \phi$, $e^{ds}_{\partial_n \phi}$,  
are shown in columns 6 and 8. 
The surface potential converges with a pattern of $O(N^{-1})$ as shown in column 7, 
which is faster than its normal derivative with a pattern of $O(N^{-1/2})$ as shown in column 9. 
We believe that this is due to the continuity of surface potential 
and the discontinuity of the normal derivatives  across the interface. 
We also can observe that  the GMRES iterations shown in column 10 in all the tests 
are less than or equal to four, 
which verifies that the boundary integral formulation is well-posed.

Note a back-to-back comparison between Table~\ref{tb_discretization} in this manuscript 
and Table 2 in our previous work \cite{Geng:2013b} shows improvements in convergence of 
$E^{ds}_{sol}$, $\phi$, and  $\partial_n \phi$ for the present work. 
This is due to the Galerkin scheme with Duffy's trick and the Cubature method in treating the singularity
as opposed to the collocation scheme with simply the removal of 
singular integral whenever it occurs on an element \cite{Geng:2013b}.

\subsubsection{Quadrature error}
In part 1, we noticed that the converge rate for $E^{ds}_{sol}$ is about $O(N^{-1/2})$ 
as the rate of $\partial_n \phi$, 
but is less than the $O(N^{-1})$ rate of $\phi$. 
To investigate the possible reason, 
we study the influence of the  quadrature rule the next.

We increase the order of the tensor product Gauss-Legendre rule to $2$, $3$ and $4$ 
and test their effects on the discretization error of $E^{ds}_{sol}$ 
as shown in Table~\ref{tb_quadrature}. 
Comparing with results in Table~\ref{tb_discretization}, 
using higher order quadratures improves both the convergence rate of $E^{ds}_{sol}$
and the required GMRES iterations. 
When the quadrature order is $4$, the electrostatic solvation energy $E^{ds}_{sol}$ 
converges to the exact energy at the rate $O(N^{-1})$ approximately. 

Increasing the quadrature order further
will not significantly improve the convergence rate of $E^{ds}_{sol}$,
because then the discretization error will be greater than the
quadrature error.  
Since higher quadrature requires more computational cost,   
in practice, due to the large size of the protein solvation problem, 
we will use quadrature order 1 as it shows the optimal combination or accuracy and efficiency. 

\begin{table}[htp]
\caption{\small Discretization error of solvation free energy $E^{ds}_{sol}$ 
for solving the same problem in case 1 using Gaussian quadrature orders of 2-4}
{
\small
\begin{center}
\begin{tabular}{lccccccccc}
\hline
 & \multicolumn{3}{c}{Quad. Order 2} & \multicolumn{3}{c}{Quad. Order 3} & \multicolumn{3}{c}{Quad. Order 4}
\\
\rule{0pt}{12pt}
$N$ & $E^{ds}_{sol}$ & $r^{ds}_{sol}$ & Iters & $E^{ds}_{sol}$ & $r^{ds}_{sol}$ & Iters & $E^{ds}_{sol}$ & $r^{ds}_{sol}$ & Iters\\\hline
320      &-8378.82&3.7&2 &-8369.13&3.9&2 &-8369.21&3.9&2\\
1280     &-8305.24&3.3&3 &-8298.57&3.8&2 &-8297.67&4.0&2\\
5120     &-8284.01&3.0&3 &-8280.12&3.7&3 &-8279.40&3.9&3\\
20,480   &-8277.27&2.7&3 &-8275.21&3.6&3 &-8374.81&4.0&3\\
81,920   &-8274.94&2.4&3 &-8273.89&3.3&3 &-8373.68&4.1&3\\
327,680  &-8274.03&2.3&3 &-8273.51&3.0&3 &-8373.40&4.0&3\\
1,310,720&-8273.64&2.2&3 &-8273.38&2.7&3 &-8273.33&4.7&3\\
$\infty$ &-8273.31&   &  &-8273.31&   &  &-8273.31&   & \\
\hline
\end{tabular}
\end{center}
}
\label{tb_quadrature}
\end{table}%

\subsubsection{FMM}
\begin{figure}[htb!]
	\begin{center}	
		\includegraphics[width=2.188in]{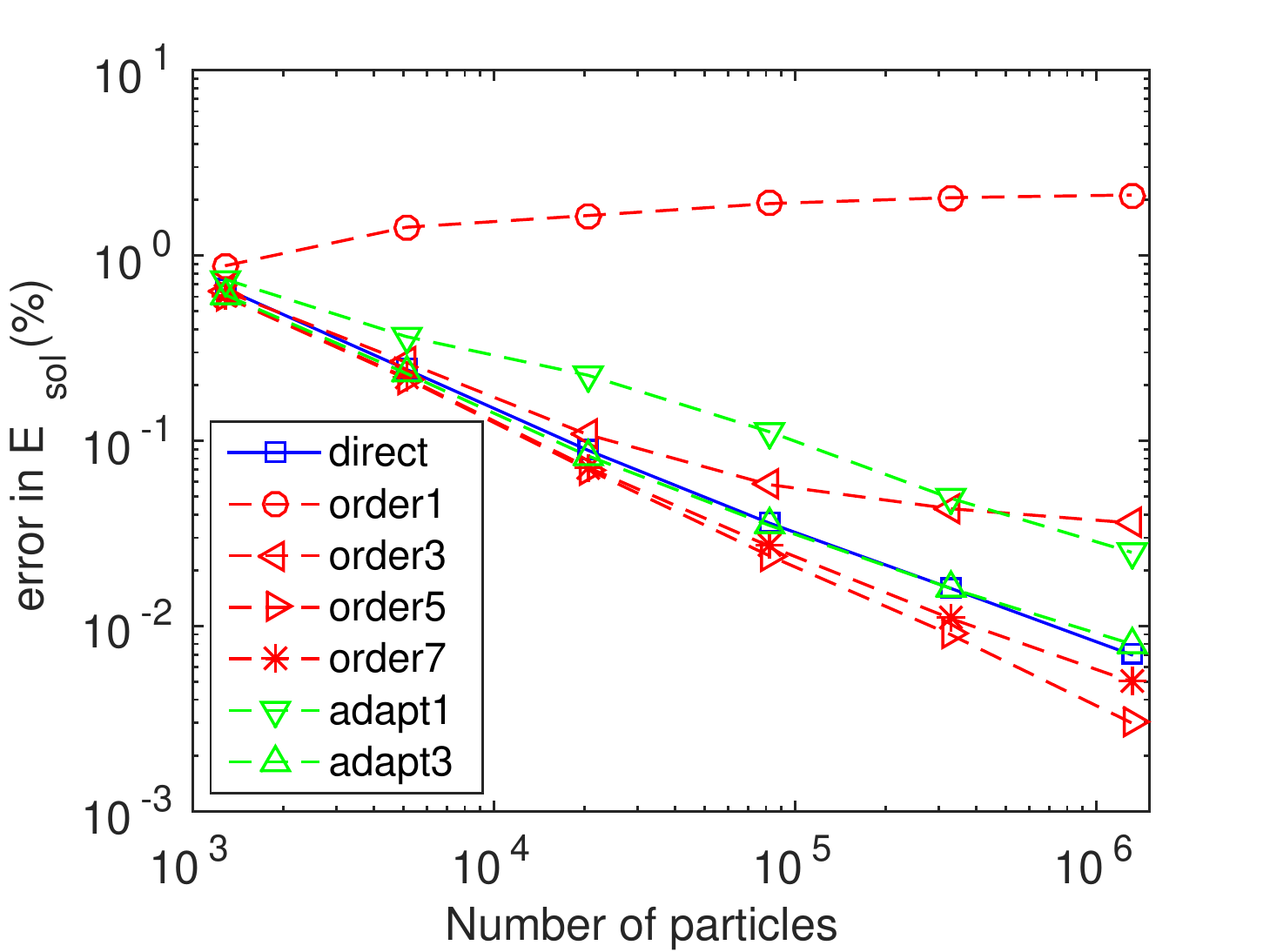}
		\includegraphics[width=2.188in]{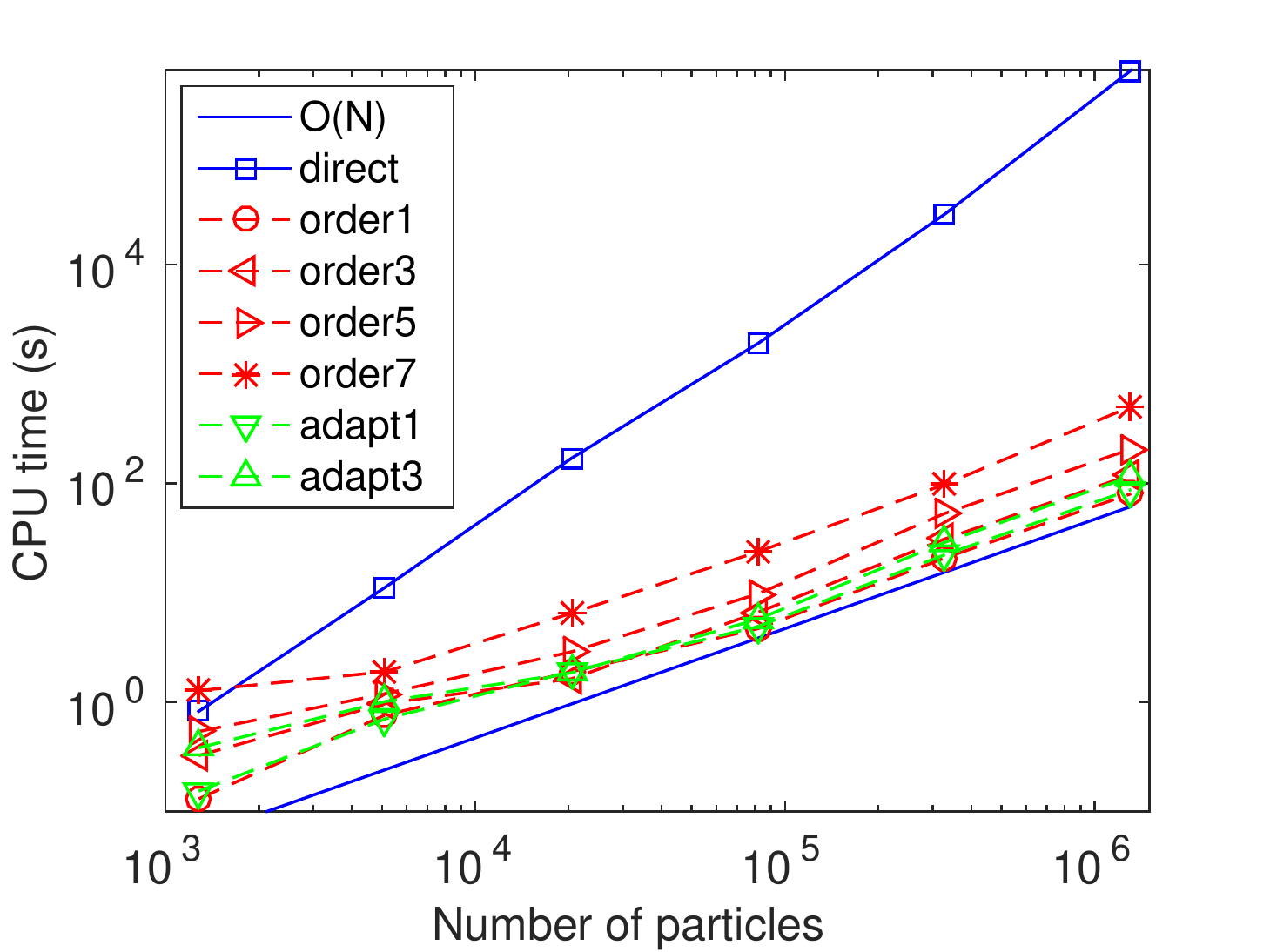}
		\includegraphics[width=2.188in]{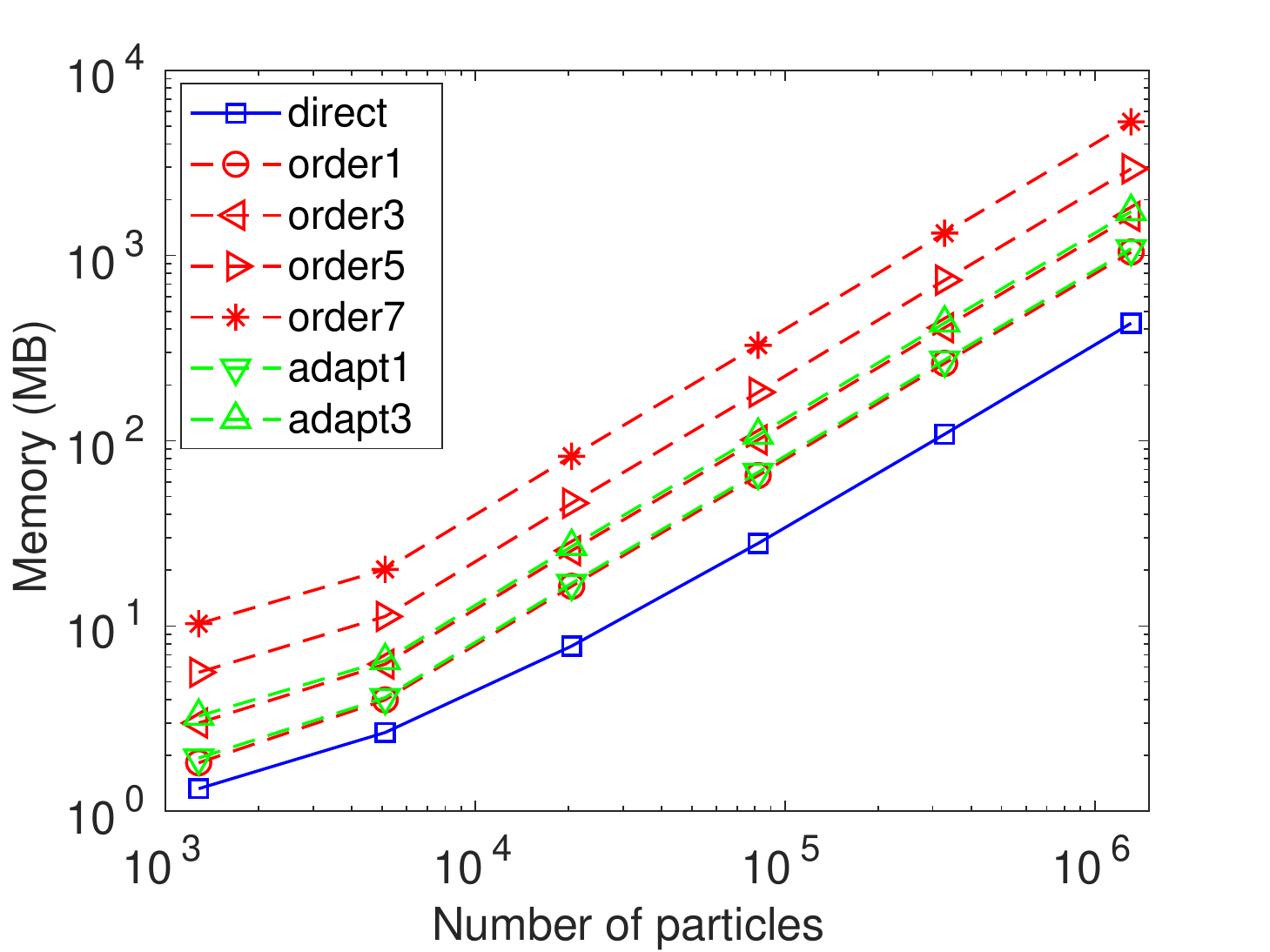}\\
		(a) ~~~~~~~~~~~~~~~~~~~~~~~~~~~~~~~~~~~~~~~~ (b)  ~~~~~~~~~~~~~~~~~~~~~~~~~~~~~~~~~~~~~ (c)
		\caption{(Compute electrostatic solvation energy on a Kirkwood sphere as number of triangles/particles $N$ in creases:  (a) Error, (b) CPU time, and (c) Memory usage; discretization error $e^{ds}_{sol}$ (solid line), Cartesian FMM approximation error $e^{cf}_{sol}$ (dashed line); Taylor expansion order $p = 1, 3, 5, 7$ and adaptive Taylor order $p=1,3$}
		\label{fig_sphere_err}
	\end{center}
\end{figure}

This part of the test studies the role of Cartesian FMM 
relating to the accuracy and efficiency of the algorithm. 
We applied the FMM to replace the direct-sum for accelerating the matrix-product calculation in GMRES. 
Here we use the first order quadrature rule for simplicity.
We set $\eta = 0.8$, which is defined in (\ref{def:eta}) and adjust
{the number of levels $L$ in the FMM algorithm} for different $N$.
Figure~\ref{fig_sphere_err} shows (a) the error in electrostatic solvation energy, (b) the CPU time, and (c) the memory usage 
versus the number of triangles $N$.
Here the error is computed as compared with the exact value $E_{sol}^{ex} = -8273.31$. 
We provide results using fixed Taylor expansion order $p=1,3,5,7$ 
and adaptive order start from $p=1$, and $p=3$. 
Here the adaptive order represents the idea that 
expansion order should be adjusted to the level (e.g. higher expansion order at higher level) 
in order to match the discretization error \cite{Tausch:2003}. 
In this figure, the solid blue line with square marks is results of direct summation with one point quadrature, 
which shows in (a) an $O(N^{-1/2})$ order of convergence in accuracy as observed in Table~\ref{tb_discretization},
an $O(N^2)$ CPU time in (b), and an $O(N)$ memory usage in (c). 

As seen in Fig.~~\ref{fig_sphere_err}(a), 
the use of FMM introduces truncation error in addition to the discretization error. 
Truncation errors are more significant than the discretization error when the order $p$ is small and are less significant when $p$ is large.

Furthermore, we observed that when expansion orders $p=5,7$ of the FMM are used, 
the errors are even smaller than those obtained with the direct sum. 
This is due to the fact that the error of the truncation error of
Taylor approximation is smaller than the quadrature error of the far
field coefficients in the direct sum.

As seen in Fig.~\ref{fig_sphere_err}(b),  the use of FMM significantly reduce the CPU time, 
which shows a $O(N)$ pattern as opposed to the $O(N^2)$ pattern of the direct sum. 
Figures.~\ref{fig_sphere_err}(a-b) combined also justify the use of adaptive order. 
Adaptive order 1 and 3 use about the same amount of CPU time as regular order 1 and 3 but achieved significant improvements in accuracy. 
Meanwhile, Figure~\ref{fig_sphere_err}(c) shows that FMM use additional memory in trading of efficiency. 
However, the $O(N)$ pattern of memory usage 
are well preserved at different orders with only an adjustment in a factor. 
 
In summary, from Tables~\ref{tb_discretization} and \ref{tb_quadrature}, 
we observed that the Galerkin discretization with piecewise constant basis functions 
can achieve $O(N^{-1/2})$ convergence rate with low quadrature order (e.g. 1 or 2)
and can achieve $O(N^{-1})$ convergence rate with high quadrature order (e.g. 4). 
Applying FMM algorithm for acceleration significantly reduced the $O(N^2)$ CPU time to $O(N)$ 
while maintains desired accuracy and $O(N)$ memory usage. 
For later tests, we apply the adaptive FMM with starting order $1$ 
and $\eta=0.8$ as an optimal choice at the consideration of both efficiency and accuracy.

\subsection{The protein 1a63}
In this section, we use the FAGBI solver to compute the solvation energy for protein 1A63, 
which has 2065 atoms. In computation involving proteins,
the molecular surface is triangulated by MSMS \cite{Sanner:1996},
with atom locations from the Protein Data Bank \cite{PDB}
and partial charges from the CHARMM22 force field \cite{Brooks:1983}.
MSMS has a user-specified density parameter $d$
controlling the number of vertices per $\text{\AA}^2$ in the triangulation.
MSMS constructs an irregular triangulation
which becomes smoother as $d$ increases.
The tree structure level is adjusted according to different number of particles.
The GMRES tolerance is $\tau=10^{-4}$.
These are representative parameter values chosen 
to ensure that the FMM approximation error and GMRES iteration error are smaller
than the direct sum discretization error,
and to keep efficient performance in CPU time and memory based on tests on spheres previously.

\begin{table}[htp]
\caption{(protein 1A63). FAGBI results; PB equations; showing electrostatic solvation energy $E_{sol}$, error, CPU time, memory usage; columns show MSMS density ($d$~in $\text{\AA}^{-2}$), number of triangles $N$, $E_{sol}$ values computed by direct sum (ds) and Cartesian FMM (cf), discretization error $e^{ds}_{sol}$, Cartesian FMM approximation error $e^{cf}_{sol}$ and their convergence rate $r^{ds}_{sol}$ and $r^{cf}_{sol}$; adaptive Taylor expansion order $p=1$, separate rate $\eta = 0.8$.}
{
\small
\begin{center}
\begin{tabular}{lrrrcrrcrrcrrcrr}
\hline
$d$ & \multicolumn{1}{l}{$N^a$} & \multicolumn{2}{l}{$E_{sol}$ (kcal/mol)} & & \multicolumn{2}{l}{Error $(\%)$} & & \multicolumn{2}{l}{Rate} & & \multicolumn{2}{l}{CPU (s)}  & & \multicolumn{2}{l}{Mem. (MB)} 
\\
\cline{3-4}\cline{6-7}\cline{9-10}\cline{12-13}\cline{15-16}
\rule{0pt}{13pt}
 & & $ds$ & $cf$ & & \multicolumn{1}{r}{$e^{ds}_{sol}$} & \multicolumn{1}{r}{$e^{cf}_{sol}$} & & \multicolumn{1}{r}{$r^{ds}_{sol}$} & \multicolumn{1}{r}{$r^{cf}_{sol}$} & & \multicolumn{1}{r}{$ds$} & \multicolumn{1}{r}{$cf$} & & \multicolumn{1}{r}{$ds$} & \multicolumn{1}{r}{$cf$}
\\
\hline
1 & 20,227&-2755.05&-2756.82&&16.12&16.20&&   &   &&    632&  6&& 18& 30\\
2 & 30,321&-2498.20&-2499.20&& 5.30& 5.34&&5.5&5.4&&   1135&  8&& 23& 40\\
5 & 69,969&-2412.40&-2413.02&& 1.68& 1.71&&2.7&2.7&&   5912& 17&& 66&111\\
10&132,133&-2383.09&-2382.50&& 0.45& 0.42&&4.1&4.4&& 36,530& 37&& 92&165\\
20&264,927&-2375.21&-2376.52&& 0.11& 0.17&&3.9&2.6&&149,651& 69&&249&423\\
40&536,781&-2371.52&-2372.77&& 0.04& 0.01&&2.9&7.5&&618,879&141&&359&654\\
  &${\infty}^b$&-2372.48&-2372.48&&&&&&&&&\\
\hline
\end{tabular}
\end{center}
${}^a$ Number of elements in triangulation.\\
${}^b$ This row shows the estimates of exact energy $E^{ex}_{sol}$ obtained by the parallel computing on high order quadrature method.
}
\label{tb_1a63}
\end{table}%

In Table~\ref{tb_1a63}, the first two columns give the MSMS density ($d$) and number of faces $N$ in the triangulation .
The next two columns give the electrostatic solvation energy $E_{sol}$ computed by
direct sum $(ds)$ and Cartesian FMM $(cf)$.
We use a parallel version of direct sum
to compute an estimate of the exact energy 
with high order quadrature methods.
We computed the discretization errors $e^{ds}_{sol}$ and $e^{cf}_{sol}$
on the fifth and sixth columns, 
which shows convergence rate faster than $O(N^{-1})$ as
observed for the geodesic grid triangulation of the Kirkwood
sphere in Case 1. 
The faster convergence seen here is due to
non-uniform adaptive treatment of MSMS triangulation \cite{Geng:2013b}.

A back-to-back comparison of results from direct sum ($ds$) and Cartesian FMM results $cf$
in Error, Rate, CPU time, and Memory in table~\ref{tb_1a63} provides the following conclusions. 
(1) the adoption of FMM only slightly modify the error and its convergence rate in accuracy, not even necessarily in a negative way; 
(2) Cartesian FMM dramatically reduces the  $O(N^2)$ direct sum CPU time to $O(N)$. 
%
For example, the simulation with $d = 10\text{\AA}^{-2}$
and $N=132,133$ took $36,530 \text{s} \approx 10 \text{h}$ by direct sum
and $37 \text{s} \approx 1/2 \text{min}$ by FMM; 
(3) Moreover, the memory usage shows that
both the direct sum and FMM memory usage is $O(N)$.
For the FMM, more memory is used for the moment and local coefficient storage
but this only adds a pre-factor rather than increases the growth rate.\\

\subsection{27 proteins}
\begin{table}[t!]
{\small
\caption{Convergence comparison using diagonal preconditioning (d) and block diagonal preconditioning (bd) on a set of 27 proteins; MSMS density $d=10$.}
\begin{center}
\begin{tabular}{rrr|rrr|rr|rrr}
\hline
Ind. & PDB & \# of ele. & \multicolumn{3}{c|}{$E_\text{sol}$ (kcal/mol)} &\multicolumn{2}{c|}{\# of it.} & \multicolumn{3}{c}{CPU time (s)}  \\
&&& d & bd & \text{diff.} (\%) & d & bd & d & bd & ratio \\\hline
1&1ajj&40496  &-1141.17&-1141.15&0.00&22&14&12.5&9.7&1.28\\
2&2erl&43214  & -953.43& -953.42&0.00&15&10&9.2&7.8&1.18\\
3&1cbn&44367  & -305.94& -305.94&0.00&12&11&7.4&8.3&0.88\\
4&1vii&47070  & -906.11& -906.11&0.00&16&14&10.6&11.5&0.92\\
5&1fca&47461  &-1206.46&-1206.48&0.00&16&11&10.2&8.8&1.16\\
6&1bbl&49071  & -991.21& -991.22&0.00&19&13&13.3&11.2&1.18\\
7&2pde&50518  & -829.49& -829.46&0.00&{\bf 75}&23&50.7&19.5&{\bf 2.60}\\
8&1sh1&51186  & -756.64& -756.63&0.00&{\bf \underline{100}}&21&70.7&18.2&{\bf 3.89}\\
9&1vjw&52536  &-1242.55&-1242.56&0.00&11&10&8.2&9.3&0.87\\
10&1uxc&53602 &-1145.38&-1145.38&0.00&20&13&14.7&11.9&1.23\\
11&1ptq&54256 & -877.83& -877.84&0.00&16&13&11.9&12.2&0.97\\
12&1bor&54628 & -857.28& -857.27&0.00&14&13&10.9&12.5&0.87\\
13&1fxd&54692 &-3318.18&-3318.14&0.00&10&10&7.8&9.9&0.79\\
14&1r69&57646 &-1094.86&-1094.86&0.00&13&12&10.6&12.6&0.84\\
15&1mbg&58473 &-1357.32&-1357.33&0.00&18&13&14.8&13.6&1.09\\
16&1bpi&60600 &-1309.61&-1310.02&0.03&18&12&16.2&14.5&1.11\\
17&1hpt&61164 & -816.47& -817.34&0.11&15&13&12.8&14.0&0.92\\
18&451c&79202 &-1031.74&-1031.91&0.02&27&20&30.3&28.8&1.05\\
19&1svr&88198 &-1718.97&-1718.97&0.00&15&12&21.4&21.3&1.01\\
20&1frd&81792 &-2868.29&-2867.32&0.00&14&12&18.1&17.2&1.05\\
21&1a2s&84527 &-1925.23&-1925.24&0.00&20&17&26.4&24.8&1.06\\
22&1neq&89457 &-1740.50&-1740.49&0.00&19&15&26.7&22.8&1.17\\
23&1a63&132133&-2382.50&-2382.50&0.00&21&16&41.3&36.8&1.12\\
24&1a7m&147121&-2171.13&-2172.12&0.00&{\bf 55}&21&111.2&51.4&{\bf 2.16}\\\hline
25&2go0&111615&-1968.61&-1968.65&0.00&{\bf 44}&24&67.6&43.0&1.57\\
26&1uv0&128497&-2296.43&-2296.43&0.00&{\bf 73}&25&130.7&52.6&{\bf 2.48}\\
27&4mth&123737&-2479.62&-2479.61&0.00&{\bf 36}&18&64.3&37.0&1.74\\
\hline
\end{tabular}
\end{center}
\label{tb_proteinSets}
}
\end{table}%

\begin{figure}[htb]
	\begin{center}
		\includegraphics[width=3.3in]{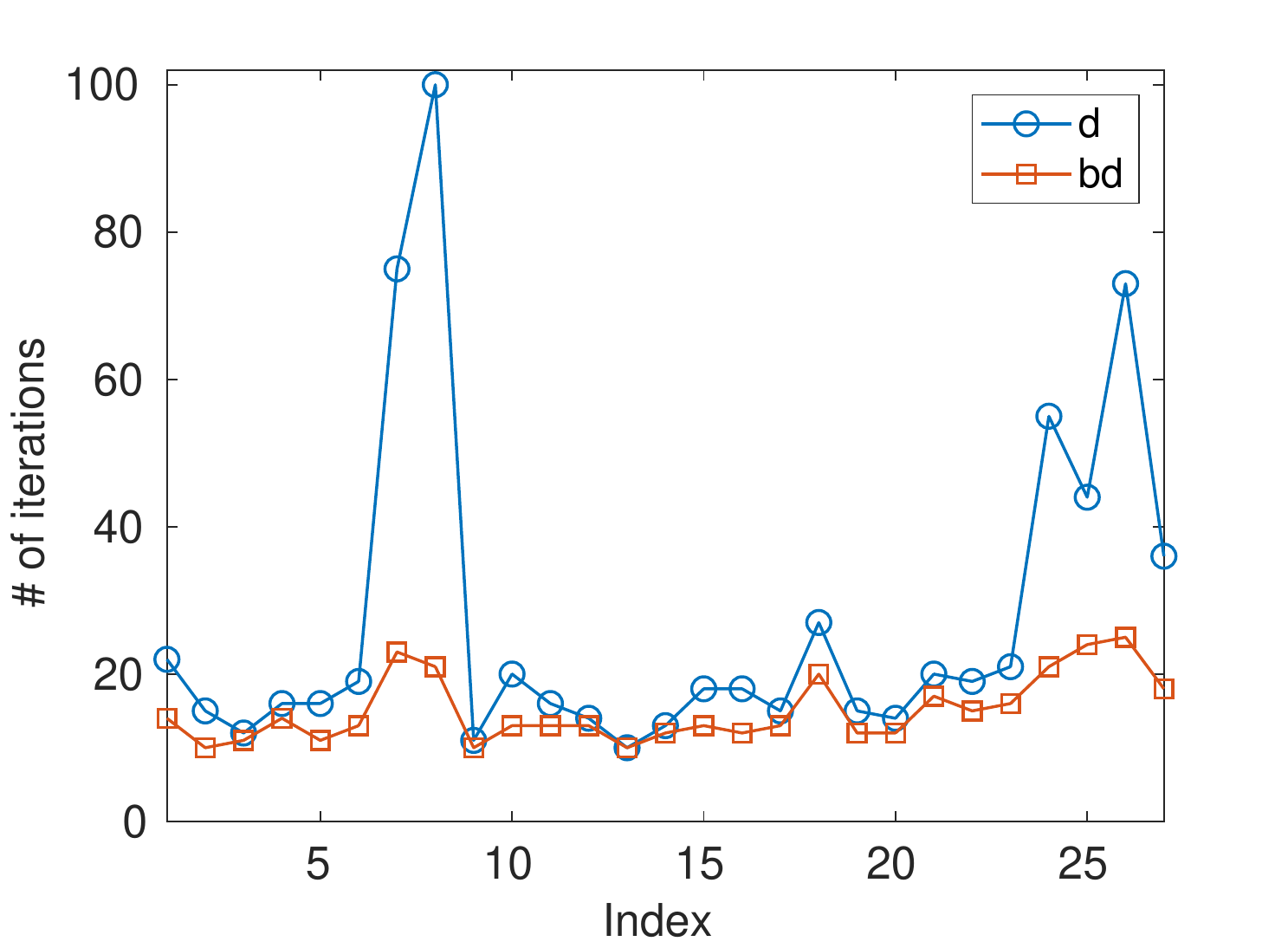}~~~~~
		\includegraphics[width=3.3in]{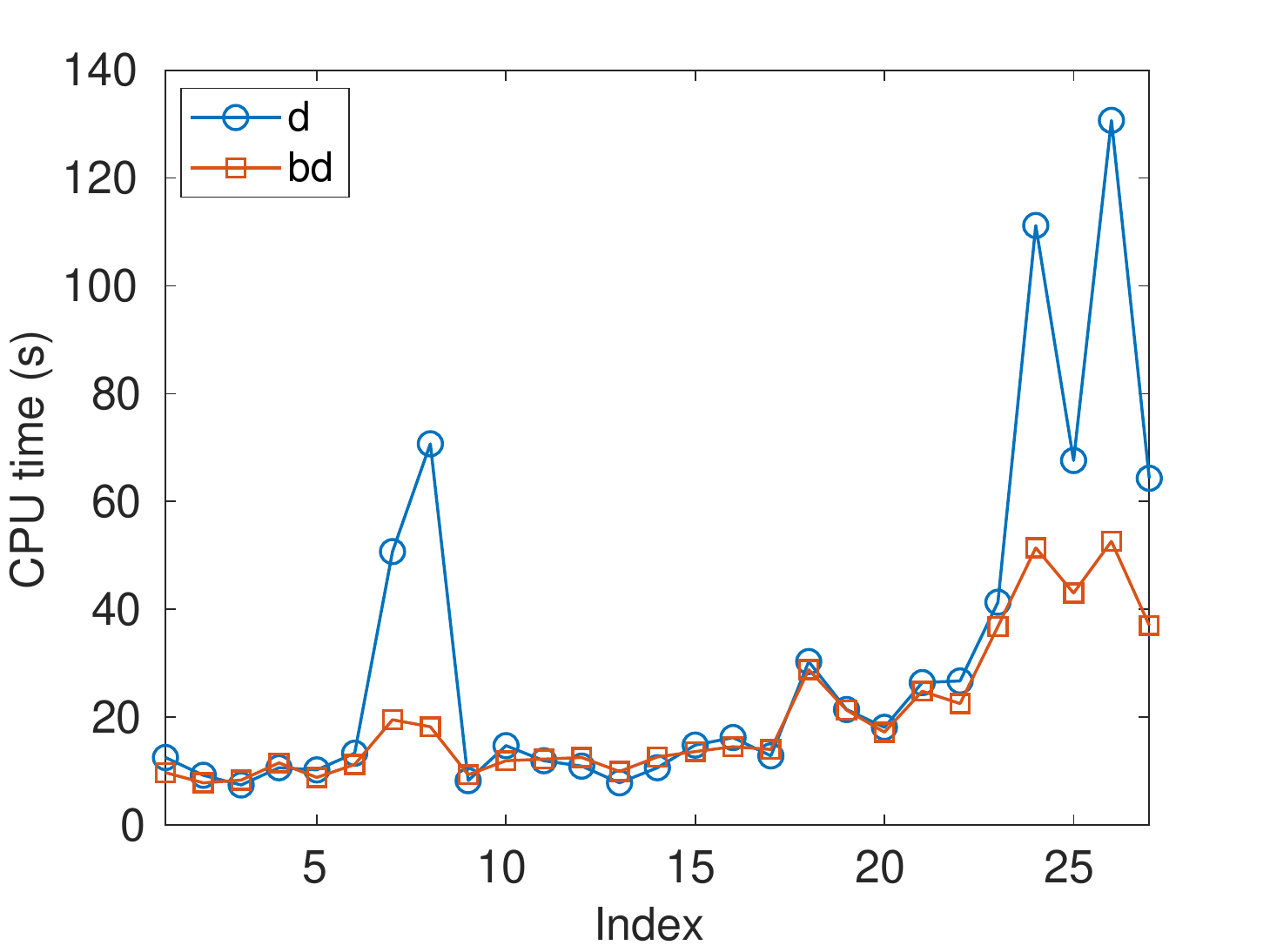}\\
		\hskip 0.3in (a) \hskip 3.25in (b)
		\caption{convergence comparison using diagonal preconditioning and block diagonal preconditioning.
			(a) number of iterations; 
			(b) CPU time (s). 
		}
		\label{fig_proteinSets}
	\end{center}
\end{figure}

We finally provide testing results on a set of 27 proteins 
for the purpose of demonstrating the general application of FAGBI solver to broader macromolecules  
and the efficiency of the preconditioning scheme. 
Table~\ref{tb_proteinSets} shows the convergence tests using diagonal preconditioning (d)
and block diagonal preconditioning (bd) for a set of 27 proteins.
After applying the block diagonal preconditioning scheme,
the cases with slow convergence using diagonal preconditioning 
has been well resolved.
In this table, the first column is the protein index,
followed by the PDB ID in the second column, 
and the number of elements in the third column generated by MSMS with density $d=10$.
Columns 4 and 5 are the solvation energy of the proteins 
applying both preconditioning schemes,
and column 6 is the relative difference between both methods, 
which shows no significant difference.
A significant reduction of number of iterations using block diagonal preconditioning (bd)
is shown in column 8 compared with results in column 7 using diagonal preconditioning (d).
One can see that the worse the diagonal preconditioning result is, 
the larger improvements block diagonal preconditioning can achieve.
For example, proteins 2pde, 1sh1, 1a7m, 2go0, 1uv0 and 4mth 
first use 75, 100, 55, 44, 73, and 36 iterations for diagonal preconditioning as highlighted in column 7, 
but only use 23, 21, 21, 24, 25, and 18 iterations for block diagonal preconditioning.
The CPU time comparison in columns 9 and 10, as well as their ratio in column 11,
further confirms the results in columns 7 and 8 
as CPU time is related to the number of iterations.
The ratio of CPU reduction for some proteins are more than 2 times as highlighted in the last column.
We plot the results of columns 7,8,9 and 10 in Fig.~\ref{fig_proteinSets}
which shows the improvements on both number of iterations
and CPU time when block diagonal preconditioning is used to replace the diagonal preconditioning.
It shows that the block diagonal preconditioning does not impair
the originally well-conditioned cases
but significantly improve the slow convergence cases, 
which suggests that we can uniformly use block diagonal preconditioning 
in replace of the original diagonal preconditioning.
Figures~\ref{fig_proteinSets}(a) and~\ref{fig_proteinSets}(b) shows a similar pattern
as CPU time and the number of iterations are highly correlated.

\section{Conclusion}
In this paper, we report recent work in developing 
an FMM accelerated Galerkin boundary integral (FAGBI) method 
for solving the Poisson-Boltzmann equation. 
The solver has combined advantages in accuracy, efficiency, and memory
as it applies a well-posed boundary integral formulation 
to circumvent many numerical difficulties
and 
uses an $O(N)$ Cartesian FMM  to accelerate the GMRES iterative solver. 
Special treatments such as
adaptive FMM order,
block diagonal preconditioning, 
Galerkin discretization,  
and Duffy's transformation are combined to improve the performance,
which is validated on benchmark Kirkwood's sphere and a series of testing proteins.  
With its attractive  $O(N^{-1})$ convergence rate in accuracy, $O(N)$ CPU run time, and $O(N)$ memory usage, the FAGBI solver and its broad usage can contribute significantly to the greater computational biophysics/biochemistry community as a powerful tool for the study of electrostatics of solvated biomolecules.

\section*{Acknowledgments}
The work of W.G. and J.C. was supported by NSF grant DMS-1418957,
DMS-1819193, SMU new faculty startup fund and SMU center for
scientific computing. The work of J.T. is in part funded by NSF grant
DMS-1720431. 


\bibliographystyle{ieeetr}
\bibliography{pb_review}
\end{document}